\def \newpage {%
  \if@noskipsec
    \ifx \@nodocument\relax
      \leavevmode
      \global \@noskipsecfalse
    \fi
  \fi
  \if@inlabel
    \leavevmode
    \global \@inlabelfalse
  \fi
  \if@nobreak \@nobreakfalse \everypar{}\fi
  \par
  \vfil
  \penalty -\@M}
\documentclass[amsmath,amsfonts,amssymb,twocolumn,superscriptaddress,showpacs]{revtex4-1}

\usepackage{graphicx,psfrag,color}
\usepackage{dcolumn}
\usepackage{bm}
\usepackage{mathbbol, appendix}
\usepackage{multirow}
\usepackage{booktabs}
\usepackage{enumitem}
\usepackage{mathrsfs}

\begin{document}
\title{Robust Topological Degeneracy of Classical Theories}
\author{Mohammad-Sadegh Vaezi}
\affiliation{Department of Physics, Washington University, St.
Louis, MO 63160, USA}
\author{Gerardo Ortiz}
\affiliation{Department of Physics, Indiana University, Bloomington,
IN 47405, USA}
\author{Zohar Nussinov}
\email{zohar@wuphys.wustl.edu}
\affiliation{Department of Physics, Washington University, St.
Louis, MO 63160, USA}
\affiliation{Department of Condensed Matter Physics, Weizmann Institute of Science, Rehovot 76100, Israel}
\date{\today}

\begin{abstract}
We challenge the hypothesis that the ground states of a physical system
whose degeneracy depends on topology must necessarily realize
topological quantum order and display non-local entanglement. To this
end, we introduce and study a classical rendition of the Toric Code
model embedded on Riemann surfaces of different genus numbers. We find
that the minimal ground state degeneracy (and those of all levels)
depends on the topology of the embedding surface alone. As the ground
states of this classical system may be distinguished by local
measurements, a characteristic of Landau orders,  this example
illustrates that topological degeneracy is not a sufficient condition
for topological quantum order. This conclusion is generic and, as shown,
it applies to many other models. We also demonstrate that certain
lattice realizations of these models, and other theories, display a
ground state entropy (and those of all levels) that is ``holographic", i.e., 
extensive in the system boundary. We find that clock and $U(1)$ gauge
theories display topological (in addition to gauge) degeneracies.
\end{abstract}

\maketitle

\section{Introduction}

The primary purpose of the current paper is to show that, as a matter of
principle, contrary to discerning lore that  is realized in many
fascinating systems, e.g., \cite{Wen,wen0,Fault-tolerant},  the
appearance of a {\it topological} ground state degeneracy {\it does not}
imply that these degenerate states are ``topologically ordered'', in the
sense that local perturbations can be detected without  destroying the
encoded quantum information \cite{Terhal}.  Towards this end, we
introduce  various models, including a classical version of Kitaev's
Toric Code   \cite{Fault-tolerant}, that  exhibit robust genus dependent
degeneracies but are nonetheless Landau ordered.  Those models do not
harbor long-range entangled ground states that cannot be told apart from
one another by local measurements. Rather, they (as well as all other
eigenstates) are trivial classical states. Along the way we will
discover that these two-dimensional classical models (including rather
mundane clock and $U(1)$ gauge like theories with four spin
interactions (specifically, Toric Clock and $U(1)$ theories that we 
will define) may not only have genus dependent symmetries and
degeneracies but, for various lattice types, may also exhibit {\it
holographic} degeneracies that scale exponentially in the system
perimeter.  Similar degeneracies also appear in classical systems having
two spin interactions. Thus, the classical degeneracies that we find may
be viewed as analogs of those in quantum models such as the Haah Code
model on the simple cubic lattice \cite{Haah,Haah1,Haah2}, a  nontrivial
theory with eight spin interactions that is topologically quantum
ordered,  and other quantum systems. To put our results in a broader
context, we first succinctly review current basic notions concerning the
different possible types of order. 

The celebrated symmetry-breaking paradigm \cite{Landau',Landau} has seen
monumental success across disparate arenas of physics. Its traditional
textbook applications include liquid to solid transitions, magnetism,
and superconductivity to name only a few examples out of a very vast
array. Within this paradigm, distinct thermodynamic phases are
associated with local observables known as order parameter(s).  In the
symmetric phase(s), these order parameters must vanish. However, when
symmetries are lifted, the order parameter may become non-zero. Phase
transitions occur at these symmetry breaking points at which the order
parameter becomes non-zero (either continuously or discontinuously).
Landau \cite{Landau} turned these ideas into a potent phenomenological
prescription. Indeed, long before the microscopic theory of
superconductivity \cite{BCS}, Ginzburg and Landau  \cite{GL} wrote down
a phenomenological free energy form in the hitherto unknown complex
order parameter with the aid of which predictions may be made. Albeit
its numerous triumphs, the symmetry-breaking paradigm might not directly
account for transitions in which symmetry breaking cannot occur. Pivotal
examples are afforded by gauge theories of the fundamental forces and
very insightful abstracted simplified renditions capturing their
quintessential character, e.g., \cite{wegner}. Elitzur's theorem
\cite{elitzur} prohibits symmetry breaking in gauge theories. Another
notable example where the symmetry breaking paradigm cannot be directly
applied is that of the Berezinskii-Kosterlitz-Thouless transition
\cite{BKT} in two-dimensional systems with a global $U(1)$ symmetry. By
the Mermin-Wagner-Hohenberg-Coleman theorem and its extensions
\cite{MW1,MW2,MW3,MW4}, such continuous symmetries cannot be
spontaneously broken in very general two-dimensional systems. 

Augmenting these examples, penetrating work illustrated that something
intriguing may happen when the quantum nature of the theory is of a
defining nature \cite{Wen}.  In particular, strikingly rich behavior was
found in Fractional Quantum Hall (FQH) systems \cite{Wen,Tsui,
Laughlin,Fradkin}, chiral spin liquids
\cite{Wen,chiral_spin_liquid,Fradkin},  a plethora of exactly solvable
models, e.g., \cite{Fault-tolerant,Anyons,kitaev_review,Levin-Wen}, and
other systems. One curious characteristic highlighted in \cite{Wen}
concerns the number of degenerate ground states in FQH fluids
\cite{wen'''}, chiral spin liquids \cite{wen',wen''}, and other
systems.  Namely, in these theories, the ground state ({\sf g.s.})
degeneracy is set by the topology alone. For instance, regardless of
general perturbations (including impurities that may break all the
symmetries of the Hamiltonian), when placed on a manifold of genus
number $g$ (the determining topological characteristic),  the FQH liquid
at a Laughlin  type filling of $\nu =1/q$ (with $q \ge 3$ an odd
integer)  universally has 
\begin{eqnarray}
\label{top_deg}
n^{\sf Laughlin}_{\sf g.s.} = q^{g}
\end{eqnarray}
orthogonal ground states \cite{wen'''}.  {\it Equation (\ref{top_deg})
constitutes one of the best known realization of topological
degeneracy.} Exact similarity transformations connect the second
quantized FQH systems of equal filling when these are placed on
different surfaces sharing the same genus \cite{FQHE_us}.  Making use of
the archetypal topological quantum phenomenon, the Aharonov-Bohm effect
\cite{AB}, it was argued that, when charge is quantized in units of
$(1/q)$ (as it is for Laughlin states),  the minimal ground state
degeneracy is given by the righthand side of Eq. (\ref{top_deg}) 
\cite{oshikawa}.  This may appear esoteric since realizing  FQH states
on Riemann  surfaces is seemingly not  feasible in the lab. Recent work
\cite{sagi} proposed the use of an  annular
superconductor-insulator-superconductor Josephson junction in which the
insulator is (an electron-hole double layer) in a FQH state (of an
identical filling) for which this degeneracy is not mathematical fiction
but might be experimentally addressed. Associated fractional Josephson
effects of this type in parafermionic systems were advanced in
\cite{Fock}.  

Historically, the robust topological degeneracy of Eq. (\ref{top_deg})
for FQH systems and its counterparts in chiral spin liquids  suggested
that such a degeneracy  may imply the existence of a novel sort of order
--- ``topological quantum order'' present in Kitaev's  Toric Code model
\cite{Fault-tolerant}, Haah's code \cite{Haah,Haah1}, and numerous other
quantum systems \cite{wen''',wen',wen'',wen2} --- a quantum order for
which no local Landau order parameter exists. As we will later review
and make precise (see Eq. (\ref{TQO})), in topologically ordered
systems, no local measurement may provide useful information.  

As it is of greater pertinence to a model analyzed in the current work,
we note that similar to Eq. (\ref{top_deg}), on a surface of genus $g$ 
the ground state degeneracy of Kitaev's Toric Code model
\cite{Fault-tolerant},  an example of an Abelian quantum double model
representing quantum error correcting  codes   (solvable both in the
ground state sector \cite{Fault-tolerant} as well as at all temperatures
\cite{symmetry1,symmetry2,fragility}), is
\begin{eqnarray}
\label{TCD}
n^{\sf Toric-Code}_{\sf g.s.} = 4^{g}.
\end{eqnarray}
Thus, for instance, on a torus ($g=1$), the model exhibits 4 ground
states while the system has a unique ground state on a topologically
trivial ($g=0$) surface with boundaries. By virtue of a simple mapping
\cite{symmetry1,symmetry2,fragility}, it may be readily established that
an identical degeneracy appears for all excited states;  that is the
degeneracy of each energy level is an integer multiple of $4^{g}$. 
Thus, the minimal degeneracy amongst all energy levels is given by
$4^{g}$. Same ground state degeneracy \cite{msb} appears in Kitaev's
honeycomb model \cite{Anyons,kitaev_review}.  As is widely known, an
identical situation occurs in the quantum dimer model
\cite{QDM,symmetry1,symmetry2}. Invoking the well-known ``$n-$ality''
considerations of $SU(n)$, leading to a basic spin of 1/2 in $SU(2)$ and
a minimal quark charge of 1/3 in $SU(3)$, it  was suggested
\cite{symmetry1,symmetry2} that in many systems, fractional charges
(quantized in units of $1/n$) are a trivial consequence of the
$\mathbb{Z}_{n}$ phase group center structure of a system endowed with 
an $SU(n)$ symmetry, which is associated with  the $n$ states comprising
the ground  state manifold. This $n$-ality type phase factors and other
considerations, prompted Sato \cite{sato} to suggest the use of
topological degeneracy (akin to that of  Eqs. (\ref{top_deg}) and 
(\ref{TCD})) as a theoretical diagnosis delineating the boundary between
the confined and the topological deconfined phases of QCD in the
presence of dynamical quarks. Other notable examples include, e.g., the
BF action for superconductors (carefully argued to not support a local
order parameter \cite{vadim}).

References \cite{symmetry1,symmetry2} examined the links between various
concepts surrounding topological order with a focus on the absence of
local order parameters. In particular,  building on a generalization of
Elitzur's theorem \cite{gelitzur,holography} it was shown how to
construct and classify theories for which no local order parameter
exists both at zero and at positive temperatures; this extension of
Elitzur's theorem unifies the treatment of classical systems, such as
gauge and Berezinskii-Kosterlitz-Thouless type theories  in  arbitrary
number of space (or spacetime) dimensions, to topologically ordered
systems. Moreover, it was  demonstrated that a sufficient condition for
the existence of  topological quantum order is the explicit presence, or
emergence, of  symmetries of {\it dimension $d$ lower} than the system's
dimension $D$,  dubbed $d$-dimensional gauge-like symmetries, and which
lead to the  phenomenon of {\it dimensional reduction}.  The
topologically  ordered ground states are connected by these
low-dimensional operator  symmetries  \cite{symmetry1,symmetry2}.  All
known examples of systems displaying topological quantum order  host
these low dimensional symmetries, thus providing a unifying framework
and organizing principle for such an order. 

As underscored by numerous pioneers, features such as fractionalization
and quasiparticle statistics, e.g.,
\cite{Wen,Laughlin,Fault-tolerant,Anyons,qps,brav,fraction1,fraction2,
fraction3,fraction4,fraction5,fraction6,fraction7,nayak,alicea}, edge
states \cite{Fault-tolerant,Anyons,nayak,edge,wedge}, nontrivial
entanglement \cite{symmetry1,symmetry2,etqo}, and other fascinating 
properties seem to relate with the absence of local order parameters and
permeate topological quantum order. While all of the above features
appear and complement the topological degeneracies found in, e.g., the
FQH  (Eq. (\ref{top_deg})), the Toric Code (Eq. (\ref{TCD})), and
numerous other systems, it is not at all obvious that one property (say,
a topological degeneracy such as those of Eqs. (\ref{top_deg}) and
(\ref{TCD})) implies another attribute (for instance, the absence of
meaningful local observables). The current work will indeed precisely
establish  the absence of such a rigid connection between these two
concepts (viz., topological degeneracy is not at odds with the existence
of a local order parameter).  

We will employ the lack of local order parameters (or, equivalently, an
associated robustness to local perturbations) as the defining feature of
topological quantum order \cite{symmetry1,symmetry2,fragility}. This
robustness condition implies  that local errors can be detected, and
thus corrected, without spoiling  the potentially encoded quantum
information. To set the stage, in what follows, we consider  a set of
$n_{\sf g.s.}$ orthonormal ground states$\{|g_\alpha
\rangle\}_{\alpha=1}^{n_{\sf g.s.}}$ with a spectral gap to all other
(excited) states. Specifically \cite{symmetry1,symmetry2}, a system will
be said to exhibit topological order at  zero temperature if and only if
for {\it{any quasi-local operator}} $\cal V$,
 \begin{eqnarray}
 \label{TQO}
 \langle g_{\alpha}|{\cal V}|g_{\beta} \rangle = v \  \delta_{\alpha,\beta} + c,
 \end{eqnarray}
where $v$ is a constant, independent of $\alpha$ and $\beta$,  and $c$
is a correction that is either zero or vanishes (typically exponentially
in the system size) in the thermodynamic limit. The physical content of
Eq. (\ref{TQO}) is clear: no possible quantity $\cal V$ may serve as an
order parameter to differentiate between the different ground states in
the ``algebraic language'' \cite{JW} where $\cal V$ is local 
\cite{symmetry1,symmetry2,holographic}. That is, all ground states look
identical locally. Similarly,  no local operator $\cal V$ may link
different orthogonal states -- the ground  states are immune to all
local perturbations.  Notice the importance of the  physical, and
consequently mathematical, language to establish topological order: A
physical system may be topologically ordered in a given language  but
its dual (that is isospectral) is not
\cite{symmetry1,symmetry2,holographic}.

Expressed in terms of the simple equations that we discussed thus far, the
goal of this work is to introduce systems for which the ground state
sector has a  genus dependent degeneracy (as in Eqs. (\ref{top_deg}) and
(\ref{TCD})) while, nevertheless, certain local observables (or order
parameters) $\cal V$ will be able to distinguish between different
ground states (thus violating Eq. (\ref{TQO})). Moreover, they  will be
connected by global symmetry operators as  opposed to low-dimensional
ones. Our conclusions are generic and, as shown, they apply to many 
classical models. The paradigmatic counterexample that we will introduce  is
a new classical version of Kitaev's Toric Code model
\cite{Fault-tolerant}.

We now turn to the outline of the paper. 
In Section \ref{GENERAL}, we generalize the standard (quantum) Toric Code model.
After a brief review and analysis of the ground states of Kitaev's Toric Code model (Section \ref{toric_quantum}),
we exclusively study our classical systems. In Section \ref{toric_classical}, we extensively study the ground states
of the classical variant of the model for different square lattices on Riemann surfaces of varying genus numbers 
$g \ge 1$.  A principal result will be that this and many other {\it classical systems exhibit 
a topological degeneracy}. We will demonstrate that an intriguing holographic degeneracy 
may appear on lattices of a certain type. As will be explained, topological as well as exponentially large in
system linear size (``holographic'') degeneracies can appear in 
numerous systems, not only in this new classical version of Kitaev's
Toric Code model \cite{Noteholo}. We further study the effect of lattice defects.  
The partition function of the 
classical Toric Code model is revealed in Section \ref{CTC} and Appendix \ref{Z}. 

In Section \ref{sec:clock}, we introduce related classical clock models.
Generalizing the considerations of Section \ref{toric_classical}, we will demonstrate that these clock
models may exhibit topological or holographic degeneracies. The ensuing analysis
is richer by comparison to that of the classical Toric Code model. Towards this end, we will construct a new
framework for broadly examining degeneracies. We then derive lower 
bounds on the degeneracy that are in agreement with 
our numerical analysis. These bounds
are {\it not confined to the ground state sector}. That is, all levels may exhibit 
topological degeneracies (as they do in the classical Toric Code model (Section \ref{CTC})).

In Section \ref{u1sec}, we will relate our results to $U(1)$ models and to 
$U(1)$ lattice gauge theories in particular.  The fact that simple 
 lattice gauge systems, that constitute a limiting case of our more general studied models, 
 such as the conventional classical Clock and $U(1)$ lattice gauge 
 theories {\it on general Riemann surfaces} (and their Toric Code extensions), 
 {\it exhibit topological}  (or, in some cases, holographic) 
 {\it degeneracies} seems to have been overlooked until now. 
In Section \ref{lattice_type}, we will study honeycomb 
and triangular lattice systems embedded on surfaces of different genus. 
In Section \ref{MM}, we will discuss yet three more regular lattice classical systems that 
exhibit holographic degeneracies. We summarize our main message and findings  
in Section \ref{conclusionss}.

 Before embarking on the specifics of these various models, we 
briefly highlight the organizing principle behind the existence of degeneracies in our 
theories. Irrespective of the magnitude and precise form of the interactions in these theories, 
the number of independent constraints between the individual interaction terms sets the 
system degeneracy. As such, the degeneracies that we find are, generally, {\it not a 
consequence of any particular fine-tuning}.

\section{The general Toric Code Model}
\label{GENERAL}
 
 We start with a general description of a class of two-dimensional 
stabilizer models defined on lattices embedded on closed manifolds with 
arbitrary genus number $g$ (the number of handles or, equivalently,  the
number of holes). The genus of a closed orientable surface  is related
to a topological invariant known as  Euler characteristic
 \begin{eqnarray}
 \label{c22}
 \chi  = 2-2g  ,
 \end{eqnarray}
which, for a general tessellation of that surface, satisfies the 
(Euler) relation
 \begin{eqnarray}
\label{Euler}
 \chi = V-E+F  .
 \end{eqnarray}
In Eq. (\ref{Euler}), $V$ is the number of vertices in the closed
tessellating polyhedron, or graph, $E$ is the number of edges, and $F$ 
the number of polygonal faces.  Assume that on each of the  $E$ edges of
the graph there is a spin $S$ degree of freedom, defining  a local
Hilbert space of size $\dim {\cal{H}} = {\sf{d}}_{Q}$, and that on each
of the $V$ vertices and $F$ faces we will have a number of  conditions
to be satisfied by the ground states of a model that we define next. 
 
\begin{figure}[htb]
\centering
\includegraphics[width=0.8 \columnwidth]{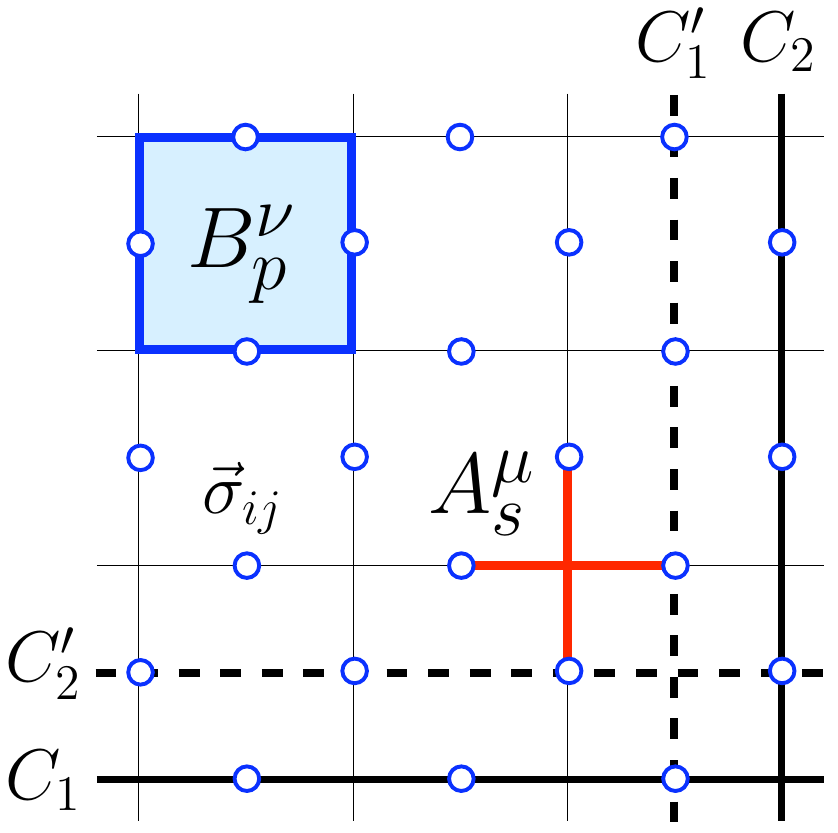}
\caption{General Toric Code lattice model with spins $S=1/2$ placed on
the edges (bonds). The red cross-shape object corresponds to the star
operator $A^\mu_s$. The plaquette operator $B^\nu_p$ is depicted in the
top-left corner in blue color. Dark solid and dashed lines represent the
loops  $C_1$, $C_2$ and $C'_1$, $C'_2$,  defining the symmetry operators
$Z_1$,  $Z_2$, and $X_1$,  $X_2$, respectively.}
\label{Sym.png}
\end{figure}

We now explicitly define, on a general lattice or graph $\Lambda$, the 
 ``General Toric Code model''. Towards this end, we consider the
 Hamiltonian 
 \begin{eqnarray}
\label{H'}
H^{\mu,\nu}= -J \sum_{s} A^{\mu}_{s} -J' \sum_{p} B^{\nu}_{p},
\end{eqnarray}
where $J$ and $J'$ are coupling constants (although it is immaterial, in
the remainder of this work we will assume these to be positive). The
interaction terms of edges in Eq. (\ref{H'}) are so-called ``star''
(``$s$'') terms ($A^\mu_{s}$) associated with the $V$ vertices (labelled by 
the letter $i$) and the $F$ ``plaquette'' (``$p$'') terms ($B^\nu_{p}$). In
the $S=1/2$ case, these are given  by the following products of Pauli
operators $\sigma^\mu_{ij}$, $\mu,\nu=x,y,z$, 
\begin{eqnarray}
\label{AB}
A^{\mu}_{s} &&= \prod_{i \in \,{\sf vertex}(s)} \sigma_{is}^{\mu} , \nonumber
\\ B^{\nu}_{p} &&= \prod_{(ij) \in \, {\sf face}(p)} \sigma_{ij}^{\nu}.
\end{eqnarray}
The product defining $A^{\mu}_{s}$ spans the spins on all edges $(is)$
that have vertex $s$ as an endpoint, and the plaquette product
$B^{\nu}_p$ is over all spins lying on the edges $(ij)$ that form the
plaquette $p$ (see Fig. \ref{Sym.png} for an illustration). A key
feature of this system (both the well known \cite{Fault-tolerant}
quantum variant ($\mu=x \neq \nu=z$) as well as, even more trivially,
the classical version that we introduce in this paper ($\mu = \nu=z$))
is that each of the bonds $A^{\mu}_s$ and $B^{\nu}_p$ can assume
${\sf{d}}_{Q} =2S+1=2$ independent values. Apart from global topological
constraints \cite{symmetry1,symmetry2} that we will expand on below, the
bonds $\{A^{\mu}_s\}$ and $\{B^{\nu}_p\}$ are completely independent of one
another. Not only, trivially, in the classical  but also  in the quantum
($q$) rendition of the model \cite{Fault-tolerant} all of these
operators commute with one another.  That is $\forall s,p \in \Lambda$,
\begin{eqnarray}
\label{triv1}
[A^{\mu}_s,B^{\nu}_p]= 0.
\end{eqnarray}

In the quantum version of the model, these terms commute as the products
defining the star and plaquette operators  must share an even number of
spins. As the individual  Pauli operators $\sigma^{x}$ and $\sigma^{z}$
appearing in the product of Eq. (\ref{AB}) anticommute, an  even number
of such anticommutations  trivially gives rise to the commutativity in
Eq. (\ref{triv1}). Even more simply, one observes that
\begin{eqnarray}
\label{triv2}
[A^{\mu}_s,A^{\mu}_{s'}]=  [B^{\nu}_{p}, B^{\nu}_{p'}]=0.
\end{eqnarray}
Lastly, from Eq. (\ref{AB}), it is trivially seen that 
\begin{eqnarray}
\label{triv3}
(A^{\mu}_s)^{2} = (B^{\nu}_p)^{2} =\mathbb{1}.
\end{eqnarray}

Apart from a number ($C_{g}^{\Lambda}$) of constraints, Eqs.
(\ref{triv1}),  (\ref{triv2}), and (\ref{triv3}) completely specify  all
the relations amongst  the operators of Eq. (\ref{AB}).  As we will
illustrate,  $H^{\mu,\nu}$ is a {\it  minimal model that embodies all of
the elements in Eq. (\ref{Euler})} such that its minimum degeneracy 
will only depend on the genus number $g$.  As all terms in the
Hamiltonian $H^{\mu,\nu}$ commute with one  another, the general Toric
Code model can be related quite trivially to a classical model.
Intriguingly, as may be readily established by a unitarity
transformation (a particular case of the bond-algebraic dualities
\cite{bond}),  the quantum version,  which includes Kitaev's Toric Code
model as a particular example,   on a graph having $E$ edges spanning
the surface of genus $g \ge 1$ is identical
\cite{symmetry1,symmetry2,fragility}, i.e.  {\it is isomorphic, to two
decoupled {\bf{classical}} Ising chains} (with one  of these chains
having $V$ classical Ising spins and the other chain composed of $F$
Ising spins) augmented by $2(g-1)$  decoupled single Ising spins. 
Perusing Eq. (\ref{H'}), it is clear that, if globally attainable,
within the ground state(s), $|g_\alpha \rangle$, 
\begin{eqnarray}
\label{gsk}
A^\mu_s |g_\alpha \rangle = (+1) |g_\alpha \rangle \ , \ 
B^\nu_s |g_\alpha \rangle = (+1) |g_\alpha \rangle ,
\end{eqnarray}
on all vertices $s$ and faces $p$ and, thus, the ground state energy is 
$E_0=-J V -J' F$.  The algebraic relations above  enable the realization
of Eq. (\ref{gsk}) for all $s$ and $p$.  

We now turn to the constraints that augment Eqs. (\ref{triv1}),
(\ref{triv2}), and (\ref{triv3}). For any lattice $\Lambda$ on any
closed surface of genus $g \ge 1$, there are $C^{\sf{universal}}_{g\ge
1} =2$ universal constraints given by the equalities
\begin{eqnarray}
\label{ABC}
\prod_{s} A^{\mu}_{s}=
\prod_{p} B^{\nu}_{p}=\mathbb{1}.
\end{eqnarray}
For the quantum variant \cite{Fault-tolerant} no further constraints
appear  beyond those of Eq. (\ref{ABC}) (that is, $C_{g}^{\Lambda} =2$
irrespective of the lattice $\Lambda$). By contrast, for  the classical
variant  of the theory realized on the relatively uncommon
``commensurate'' lattices, additional constraints will augment those of
Eq. (\ref{ABC}) (i.e., for classical systems, $C_{g}^{\Lambda} \ge 2$). 
Invoking the $C_{g}^{\Lambda}$ constraints as well as the trivial
algebra of Eqs. (\ref{triv1}) and  (\ref{triv2}), we  may transform from the
original variables -- the spins on each of the $E$ edges --
$\{\sigma^{\mu}_{ij}\}$ to new basic degrees of freedom -- all ${\sf
N}_{{\sf ind.~bonds}}$ independent {\it ``bonds''} $\{A^{\mu}_{s \neq s'}\},
\{B^{\nu}_{p \neq p'}\}$ that appear in the Hamiltonian and 
$N_{\sf{redundant}} = (E-{\sf N}_{{\sf ind.~bonds}})$ remaining
redundant spins of the original form $\{\sigma^{\mu}_{ij}\}$ on which
the energy does not depend (and thus relate to symmetries). If the bonds
$A^{\mu}_{s}$ and $B^{\nu}_{p}$ do not adhere to any constraint apart
from that in Eq. (\ref{ABC})  then ${\sf N}_{{\sf ind.~bonds}}= (V+F-2)$
of the $(V+F)$ bonds in the Hamiltonian of Eq. (\ref{H'}) will be
independent of one another. Correspondingly,  $N_{\sf{redundant}} =
[E-(V+F-2)]=2g$. As all bonds must satisfy the constraint of Eq.
(\ref{ABC}) and thus ${\sf N}_{{\sf ind.~bonds}} \le (V+F-2)$, the
number of redundant spin degrees of freedom $N_{\sf{redundant}}  \ge
2g$. In the general case, if there are $(C^{\Lambda}_{g}-2)$
constraints that augment the two restrictions already present in Eq.
(\ref{ABC}), then we may map the original system of $E$ spins to ${\sf
N}_{{\sf ind.~bonds}}= (V+F-C^{\Lambda}_{g})$ independent bonds in Eq.
(\ref{H'})  and $N_{\sf{redundant}} =
(E-{\sf N}_{{\sf ind.~bonds}})=2(g-1)+C_{g}^{\Lambda}$ spins that have no
impact on the energy.  Thus, for genus $g \ge 1$ surfaces, the
degeneracy of each energy level is an integer multiple of the {\it
minimal} degeneracy possible,
\begin{eqnarray}
\label{mindeg}
\min(n_{ \sf g.s.}) = 2^{N_{\sf{redundant}}} = n^{\sf min}_{\sf g.s.} \times
2^{C^{\Lambda}_{g}-2} ,
\end{eqnarray}
with $n^{\sf min}_{\sf g.s.}=4^{g}$. 
Equation (\ref{mindeg}) will lead to a global redundancy factor in the
partition function ${\cal Z} = {\sf Tr} \exp(-\beta H^{\mu,\nu})$ with
$\beta$  the inverse temperature.

We now focus on the ground state sector. If there are no constraints
apart from Eq. (\ref{ABC}), then to obtain the ground states it suffices
to make certain that ${\sf N}_{{\sf ind.~bonds}}$ of the  bonds are unity in
a given state. Once that occurs, we are guaranteed a ground
state in which each bond in the Hamiltonian of Eq. (\ref{H'}) is
maximized (i.e., Eqs. (\ref{gsk}) are satisfied). A smaller number of
bonds fixed to one will not ensure that only ground states may be
obtained. Thus the values of all ${\sf N}_{{\sf ind.~bonds}}$ independent
bonds need to be fixed in order to secure a minimal value of the energy.
The lower bound of the degeneracy  on each level (Eq.
(\ref{mindeg})) is saturated for the ground state sector where it 
becomes an equality. That is, very explicitly, the ground state 
degeneracy is given by
\begin{eqnarray}
\label{ngtc}
n^{\sf General~Toric-Code}_{\sf g.s.} =  4^{g} \times 2^{C^{\Lambda}_{g}-2}.
\end{eqnarray}
The equalities of Eqs. (\ref{mindeg}) and (\ref{ngtc}) are basic facts that
will be exploited in the present article. The degeneracy of Eq.
(\ref{ngtc}) is in accord with the general result 
\begin{eqnarray}
\label{general-deg}
 n^{\sf{g\geq1}}_{\sf g.s.}=
{\sf{d}}_{Q}^{-\chi+(C^{\Lambda}_{g}-C^{\Lambda}_{1})}
n^{\sf{g=1}}_{\sf g.s.} ,
\end{eqnarray}
and differs from that of Kitaev's Toric Code  model
\cite{Fault-tolerant} (Eq. (\ref{TCD})) by a factor of
$2^{C^{\Lambda}_{g}-2}$. As each of the $C^{\Lambda}_{g}$ constraints as
well as increase in genus number leads to a degeneracy of the spectrum,
a simple ``correspondence maxim'' follows: {\it it must be that we may
associate a corresponding independent set of symmetries with any
individual constraint}. Similarly, as Eqs. (\ref{mindeg}, \ref{ngtc})
attest, elevating the genus number $g$ must introduce further
symmetries. Thus, the global degeneracy of Eq. (\ref{mindeg}) is a
consequence of all of these symmetries.  

Given Eq. (\ref{H'}) it is readily seen that  the system has a gap of
magnitude $\Delta = 4(J +J')$ between the ground state $E_0$  and the
lowest lying excited state $E_1$. All energy levels $E_{\ell}$, 
defining the spectrum of $H^{\mu,\nu}$,  are quantized in integer
multiples of $J$ and $J'$. 

\section{Ground states of the quantum Toric Code model}
\label{toric_quantum}

In Kitaev's Toric Code model \cite{Fault-tolerant} the
symmetries associated with the constraints of Eq. (\ref{ABC}) are rather
straightforward, and cogently relate to the topology of the surface on
which the lattice  is embedded. An illustration for the square lattice
is depicted in Fig. \ref{Sym.png}. For such a model on a simple torus
(i.e., one with genus $g=1$), the four canonical symmetry operators
are  
\begin{eqnarray}
\label{zx}
Z^{q}_{1,2}= \prod_{(ij) \in C_{1,2}} \sigma_{ij}^{z} \ , \
X^{q}_{1,2}= \prod_{(ij) \in C'_{1,2}} \sigma_{ij}^{x} .
\end{eqnarray}
These two sets of non-commuting operators \cite{Fault-tolerant} 
\begin{eqnarray}
\label{CR}
\{X^{q}_1,Z^{q}_1\}= \ &0& \ =\{X^{q}_2,Z^{q}_2\} , \nonumber\\
\left[X^{q}_1,X^{q}_2\right]= \ &0& \ =\left[Z^{q}_1,Z^{q}_2\right] , \nonumber\\
\left[X^{q}_1,Z^{q}_2\right]= \ &0& \ =\left[X^{q}_2,Z^{q}_1\right] , 
\end{eqnarray}
realize a $\mathbb{Z}(2)\times \mathbb{Z}(2)$ symmetry and ensure 
a four-fold degeneracy (or, more generally a degeneracy that is an
integer multiple of four)  of the whole spectrum. 

To see this, we may, for instance, seek mutual eigenstates of the
Hamiltonian $H^{x,z}$ along with the two symmetries $Z^{q}_{1}$ and
$Z^{q}_{2}$ with which it commutes. Noting the algebraic relations amongst
the above operators, a moment's reflection reveals that a possible candidate
for a normalized ground state is 
given by
\begin{eqnarray}\hspace*{-0.3cm}
| g_1 \rangle= \frac{1}{\sqrt{2}}\prod_s \left ( \frac{\mathbb{1} +
A_s^x}{\sqrt{2}} \right ) |{\bf F}\rangle  , 
\end{eqnarray}
where $\sigma^z_{ij}   |{\bf F}\rangle= |{\bf F}\rangle$, for all $E$
edges, and  $\langle {\bf F}|{\bf F}\rangle=1$. This corresponds to
$Z^{q}_{1,2}| g_1 \rangle=| g_1 \rangle$. Now, because $X^{q}_{1,2}$ are
symmetries, by the algebraic relations of  Eq. (\ref{CR}), the three
additional orthogonal states 
\begin{eqnarray}\hspace*{-0.3cm}
| g_2 \rangle=X_1^q | g_1 \rangle \ , \  | g_3 \rangle = X_2^q | g_1 \rangle \ , \ 
| g_4 \rangle=X_1^q X_2^q | g_1 \rangle ,
\end{eqnarray}
are the remaining ground states. That is, the $C^{\Lambda}_{g=1}=2$ lattice
($\Lambda$)  independent constraints of the quantum model (Eq.
(\ref{ABC})) correspond to the $2$ sets of symmetry operators associated
with the $\gamma=1,2$ toric cycles ($\{Z^{q}_{\gamma}, X^{q}_{\gamma}\}$) of Eq.
(\ref{zx}). This correspondence is in agreement with the simple maxim
highlighted above. The symmetry operators $X^q_{1}$ and $Z^q_{1}$ are
independent (and trivially commute) with the symmetry operators
$X^q_{2}$ and $Z^q_{2}$.  Notice that in the spin $(\sigma^{\mu}_{ij}$)
language the ground states  above are entangled, and they are connected
by $d=1$ symmetry  operators \cite{symmetry1,symmetry2}. Moreover, the
anyonic statistics  of its excitations is linked to the entanglement
properties of those  ground states \cite{symmetry1,symmetry2}.  As
mentioned above, the model  can be trivially related, by duality, to two
decoupled classical Ising  chains so that in the dual language the
mapped ground states are  unentangled \cite{symmetry1,symmetry2}.

For a Riemann surface of genus $g$, we may write down trivial extensions
of Eqs. (\ref{zx}) for the $(2g)$ cycles circumnavigating the $g$
handles of that surface. That is, instead of the four operators of Eq.
(\ref{zx}), we may construct $2g$ operators pairs with each of these
pairs associated with a particular handle $h$ (where $1 \le h \le g$),
containing the four operators $\{Z^{q}_{\gamma,h}\}$ and $\{X^{q}_{\gamma,h}\}$
with $\gamma=1,2$. A generalization of Eqs. (\ref{CR}) leads to an algebra
amongst the $2g$ independent pairs of symmetry operators.  The
multiplicity of independent symmetries leads to the first factor in Eq.
(\ref{ngtc}). The number of constraints is, in the quantum case, lattice
independent and given by  $C^{\Lambda}_{g \ge 1} =2$ (there are no constraints
beyond those in Eq. (\ref{ABC})). It is rather straightforward to
establish that when $g=0$ (i.e., for topologically trivial surfaces),
the ground state of the quantum model is unique. Putting all of these
pieces together, the well known degeneracy of Eq. (\ref{TCD}) follows. 

\section{Ground states of the classical Toric Code model}
\label{toric_classical}

We now finally turn to the examination of the ground states of the
classical rendering of Eq. (\ref{H'}) in which only a single component
$\mu =\nu =z$ of all spins appears. We will explain how {\it the
degeneracy of Eqs. (\ref{mindeg}) and (\ref{ngtc}) emerges}. The upshot
of our analysis, already implicitly alluded to above, consists of two
main results: 
\bigskip

$\bullet$ In the most frequent lattice realization of this classical
model, {\it its degeneracy will still be given by Eq. (\ref{TCD})},
i.e., $4^{g}$.  That is,  in the most common of geometries, the number
of  ground states  will  depend on topology alone (i.e., the genus
number $g$ of the embedding  manifold). For arbitrary square lattice or
graph, as our considerations universally  mandate,  the {\it minimal}
possible ground state degeneracy will be given  by the topological
figure of merit of Eq. (\ref{TCD}). 
\bigskip

$\bullet$ In the remaining lattice realizations, {\it the degeneracy of
the system will typically be holographic}. That is, in these slightly
rarer lattices, the ground state degeneracy will scale as
${\cal{O}}(2^{L})$ where $L$ is the length of one of the sides of the
two-dimensional lattice.

\bigskip

As will be seen, for the square lattice, depending on the parity of the
length of the lattice sides, the number of constraints $C^{\Lambda}_{g}$
may exceed its typical value of two. This will then lead to an enhanced
degeneracy vis a vis the minimal possible value of $4^{g}$. In the next
subsection  we first broadly sketch the constraints and symmetries of
the classical system. As it will be convenient to formulate our main
result via the ``correspondence maxim'', we will then proceed to
explicitly relate the constraints and symmetries to one another. The
$symmetry \leftrightarrow constraint$ consonance, along with Eqs.
(\ref{mindeg}) and (\ref{ngtc}),  will then rationalize all of the
degeneracies  found for general square lattices embedded on Riemann
surfaces  of arbitrary genus number. Exhaustive calculations for these
degeneracies will then be reported  in the subsections that follow.

\subsection{Symmetries and constraints}
\label{sac}

We next  list the general symmetries and constraints of the classical
Toric Code model in square lattices of varying sizes. Consider first a 
lattice $\Lambda$  of size $L_{x} \times L_{y}$ on a torus (i.e., having
$V=L_{x} L_{y}$ vertices and  $E=2 L_{x} L_{y}$ edges). We will then
examine more general  lattices  of arbitrary genus $g$.  The square
lattice on the torus will be categorized as being one of two types:
\begin{eqnarray}
\label{type12}
\begin{cases}
	\text{Type I},&\text{$L_x \neq L_y$ where at least} \\
	&\text{one of $L_x$ or $L_y$ is odd} \\
	\text{Type II} ,&\text{otherwise}.\\
\end{cases}
\end{eqnarray}  

Type I lattices, as defined for the $g=1$ case above and their
generalizations for higher genus numbers $g >1$, only admit two
constraints $C^{\Lambda}_{g}$ and thus by the correspondence maxim only
two symmetries. For these lattices,  we will show that the ground state
degeneracy is $4^{g}$. By contrast, Type II lattices have a larger
wealth of constraints, $C^{\Lambda}_{g}>2$, and therefore  a larger
number of symmetries and a degeneracy higher than $4^{g}$. 

\begin{figure}[htb]
\centering
\includegraphics[width=0.8 \columnwidth]{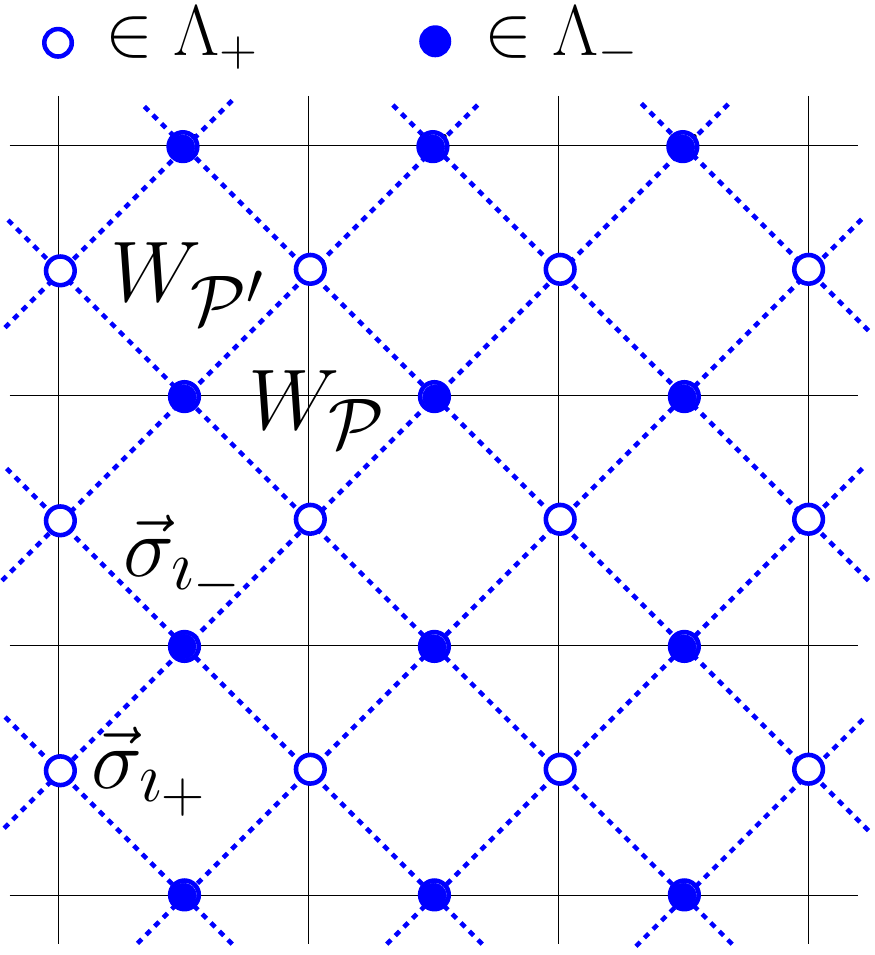}
\caption{ Dotted lines represent the rotated lattice $\Lambda'$. 
The spin degrees of freedom 
$\vec{\sigma}$ reside on the vertices of the rotated bipartite 
lattice $\Lambda'$, formed out of two sublattices $\Lambda_{+}$ 
and $\Lambda_{-}$.} 
\label{XM.png}
\end{figure}

The centers of  all nearest neighbor edges on the square lattice (of
lattice constant $a$) form yet another square lattice $\Lambda^{'}$  (of
lattice constant $a/\sqrt{2}$) at an angle of  $45^\circ$ relative to
the original lattice (Fig. \ref{XM.png}). The spins are located at the 
vertices of the rotated square lattice $\Lambda^{'}$.  In order to
describe  the symmetries and constraints of this system, let us denote
the two (standard) sublattices of the square lattice $\Lambda^{'}$ by
$\Lambda_{\pm}$. That is,  both $\Lambda_{+}$ and $\Lambda_{-}$ are, on
their own, square lattices with $\Lambda^{'} = \Lambda_{+} \cup
\Lambda_{-}$ and $\Lambda_{+} \cap \Lambda_{-} = \emptyset$.  Let us
furthermore denote the sites of $\Lambda_{\pm}$ by ${\imath_{\pm}}$,
respectively.

With these preliminaries, it is trivial to verify that 
\begin{eqnarray}
\label{t+-}
T^x_{+} = \prod_{\imath_{+}\in \Lambda_+} \sigma^{x}_{\imath_{+}},  \ \ 
\nonumber \\ 
T^x_{-} = \prod_{\imath_{-}\in \Lambda_-} \sigma^{x}_{\imath_{-}}, \ \ 
\end{eqnarray}
are, universally, both symmetries of the classical ($\mu = \nu =z$)
version of the Hamiltonian of Eq. (\ref{H'}).   Most square lattices
(those of Type I in Eq. (\ref{type12})) will only exhibit the two
symmetries of Eq. (\ref{t+-}).  The more commensurate Type II lattices
admit diagonal contours (connecting nearest neighbors of sites $\imath$
of $\Lambda^{'}$) that close on themselves before threading all of the
lattice sites of $\Lambda^{'}$. That is, in Type II lattices,  it is
possible to find  diagonal loops $\Gamma_{m}$ at a constant
$45^{\circ}$ angle (or a more non-trivial alternating contour) that
contain only a subset of all sites of $\Lambda^{'}$ (or, equivalently, a
subset of all edges $( i j )$ of the original square lattice $\Lambda$).
Associated with each such independent contour $\Gamma_{m}$, there is
a symmetry  operator,
\begin{eqnarray}
\label{tl}
T^x_{m} = \prod_{\imath \in \Gamma_{m}} \sigma^{x}_{\imath},
\end{eqnarray}
augmenting the symmetries of Eq. (\ref{t+-}). 

The form of the symmetries suggests the distinction between Type I and
Type II lattices on general surfaces. On Type II lattices, it is
possible to find, at least,  one diagonal contour $\Gamma_{m}$ that
contains a subset of all edges $( i j )$ of the lattice $\Lambda$.
Conversely, due to the lack of the requisite lattice commensurability,
on Type I lattices, it is impossible to find any such contour.  

We now turn to the constraints associated with Type I and II lattices.
These are in one-to-one correspondence with the  symmetries of Eqs.
(\ref{t+-}) and (\ref{tl}).  Specifically for Type I lattices, the only
universal constraints present are those of Eq. (\ref{ABC}) which we
rewrite again for clarity,
\begin{eqnarray}
\label{asbp}
{\cal{C}}_{+}: ~ ~ \prod_{s} A^z_{s} =1, \nonumber
\\  {\cal{C}}_{-}: ~~\prod_{p} B^z_{p} =1.
\end{eqnarray}
These two constraints  match the two symmetries of Eq. (\ref{t+-}). In
the case of the more commensurate lattices $\Lambda$, additional
constraints appear.  In order to underscore the similarities to the
symmetries of Eq. (\ref{tl}), we will now aim to briefly  use the same
notation concerning the lattice $\Lambda^{'}$. Within the framework
highlighted in earlier sections, the spin products $\{A^z_{s}\}$ and
$\{B^z_{p}\}$ of Eq. (\ref{AB}) are associated with geometrical objects
that look quite different (i.e., ``stars'' and ``plaquettes''), see Fig.
\ref{Sym.png}.  If we now label the plaquettes of $\Lambda^{'}$ by
${\mathcal{P}}$ then, we may, of course, trivially express Eq.
(\ref{H'}) as a sum of local terms, 
\begin{eqnarray}
\label{HW}
H = -J \sum_{ {\mathcal{P}}} W_{{\mathcal{P}}}-J' \sum_{ {\mathcal{P}'}} 
W_{{\mathcal{P}'}},
\end{eqnarray} 
where $W_{{\mathcal{P}}} = \prod_{\imath \in  {{\mathcal{P}}}} 
\sigma^{z}_{\imath}$ are the products of all Ising spins at sites
$\imath$ belonging to plaquette $ {{\mathcal{P}}}$.  This trivial
description renders the original star and plaquette terms of Eq.
(\ref{H'}) on a more symmetric footing, see Fig. \ref{XM.png}.

Associated with each of the symmetries of
Eq. (\ref{tl}) there is a corresponding constraint,
\begin{eqnarray}
\label{wl}
{\cal{C}}_{m}: ~~ \prod_{\imath \in \Gamma_{m}} W_{m} =1. 
\end{eqnarray}
In accordance with our earlier maxim, insofar as counting is concerned,
we have the following correspondence between the symmetries and the
associated constraints,
\begin{eqnarray}
\label{c12l}
\begin{cases}
	T^x_{+} \leftrightarrow {\cal{C}}_{+}, \\
	T^x_{-} \leftrightarrow {\cal{C}}_{-} ,\\
	T^x_{m} \leftrightarrow {\cal{C}}_{m} .\\
\end{cases}
\end{eqnarray}  
In Type I systems, wherein only the $C^\Lambda_{g}=2$ universal
constraints appear, the degeneracy of the spectrum is exactly $4^{g}$. 
In Type II lattices, $C_{g}^{\Lambda} >2$ (with the difference of
$(C_{g}^{\Lambda}-2)$ equal to the number of additional independent
contours $\Gamma_{m}$ that do not contain all edges of the original
lattice $\Lambda$)  and, as Eq. (\ref{ngtc}) dictates,  the ground state
degeneracy exceeds the minimal value of $4^{g}$ multiplied by two raised
to the power of the number of the additional independent loops.

\begin{figure}[htb]
\centering
\includegraphics[width=1.0 \columnwidth]{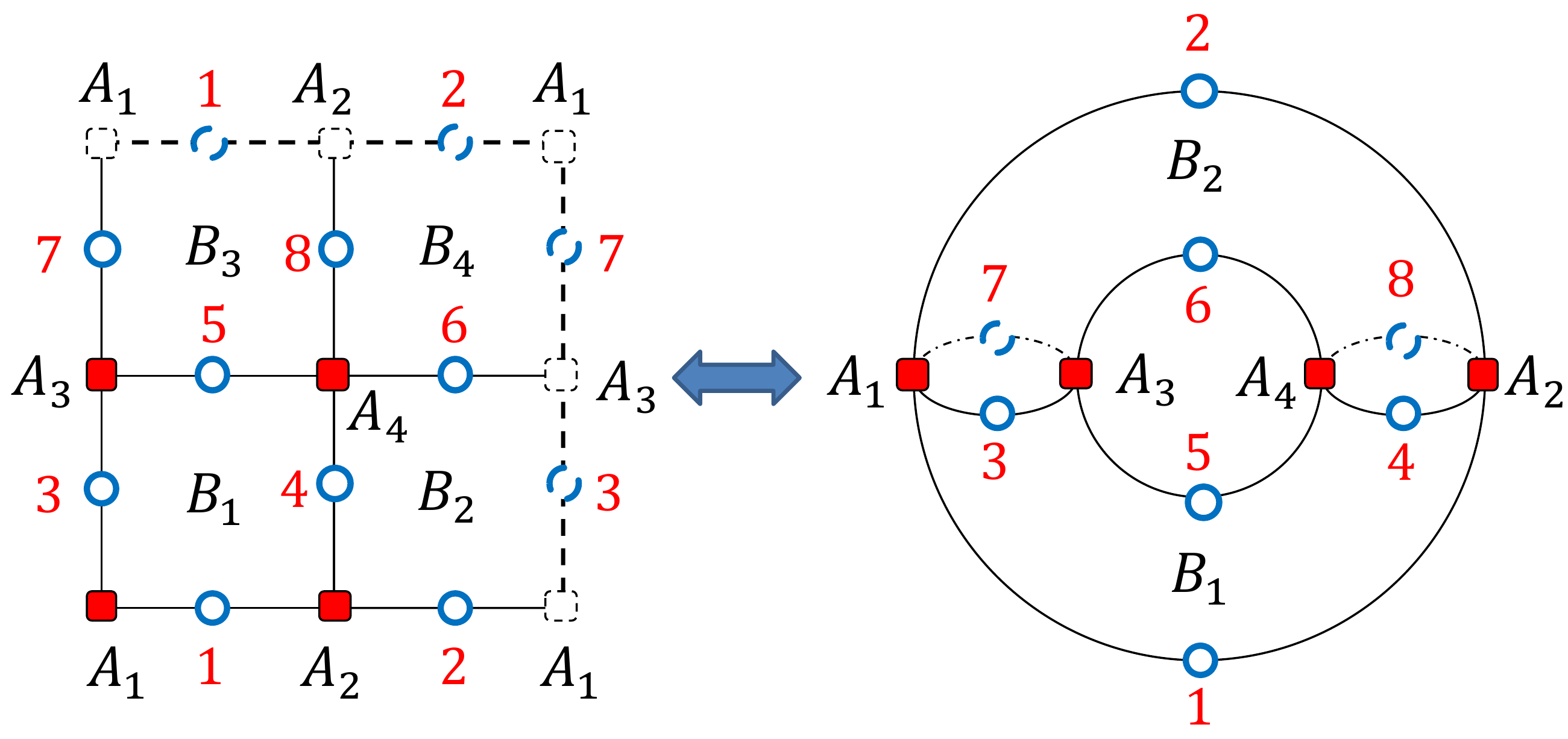}
\caption{ A square lattice with $8$ spins along with its embedding on a
torus. Because of periodic boundary conditions, spins on  boundary edges
(dashed-blue) display numbers identical to those in the  bulk. In this
figure $A_s=A^z_s$ and $B_p=B^z_p$. 
In the right panel, each edge has been labeled according to the left
panel, and the solid red squares represent the vertices labeled by
$A_{s}$. Since $B_3$ and $B_4$ are respectively behind $B_1$  and $B_2$,
we cannot see them here.}
\label{8-1.png}
\end{figure}
\subsection{Ground state degeneracy on $g=1$ surfaces}
\label{subsection CTC}

\begin{figure}[htb]
\centering
\includegraphics[width=0.91 \columnwidth]{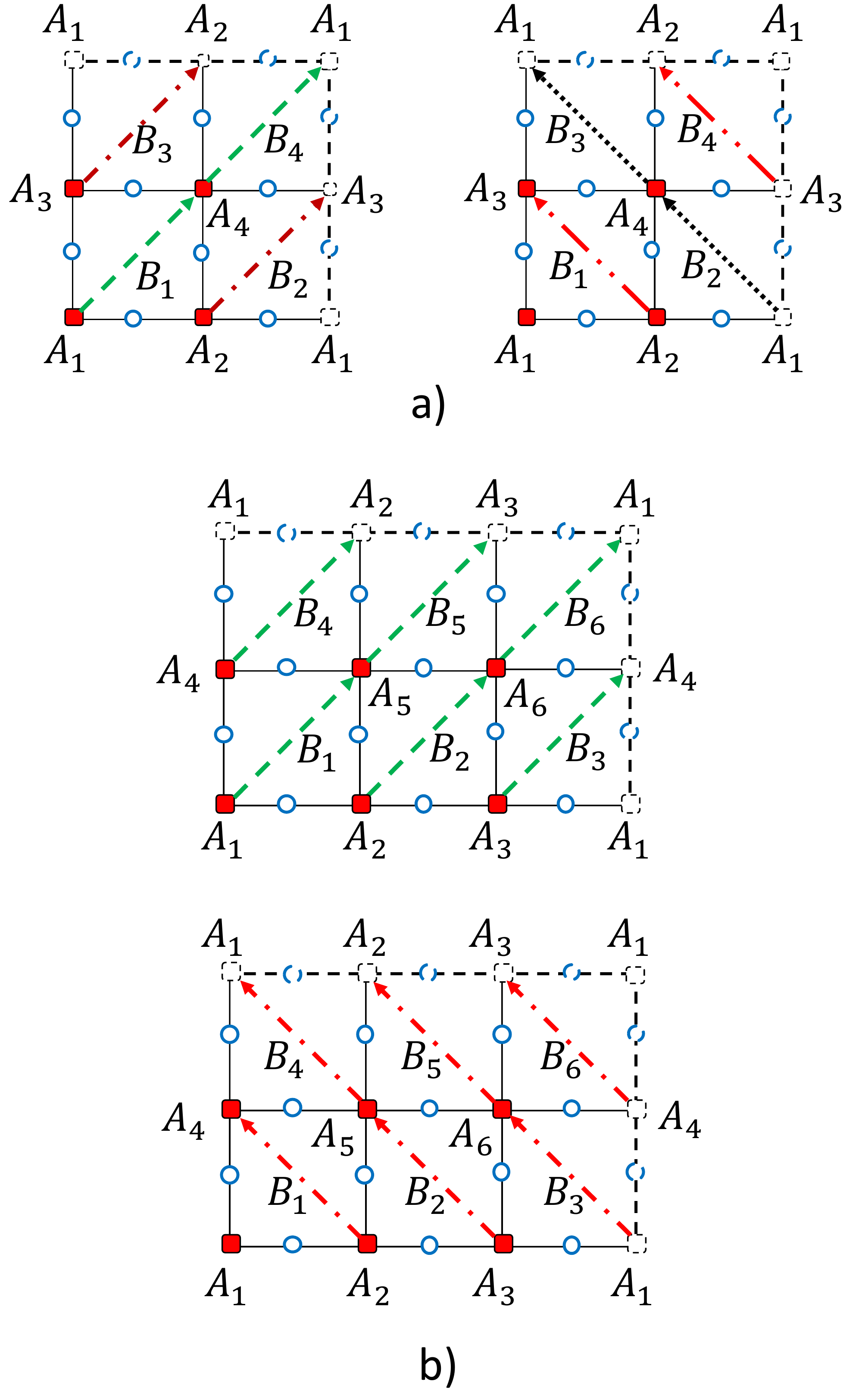}
\caption{a) Lattice of size $L_x=2$, $L_y=2$, $E=8$ and b) 
$L_x=2$, $L_y=3$, $E=12$. Diagonal lines with arrows represent possible
paths realizing constraints on $A_s=A^{z}_s$ and $B_p=B^{z}_p$.}
\label{8-2.png}
\end{figure}

Thus far, our discussion has been quite general and, admittedly,
somewhat abstract. We now turn to simple concrete examples. We first
consider the classical Toric Code model on a simple torus (i.e., a
surface with genus $g=1$), and examine small specific square
lattices of dimension $L_{x} \times L_{y}$. 
We find that for general lattices $\Lambda$ (with reference to Eq.
(\ref{type12})), the total number of independent constraints is
\begin{eqnarray}
\label{cdeg}
C^{\Lambda}_{g=1}=
\begin{cases}
	2,&\text{$\Lambda$ is a Type I lattice}\\
	&\\
	2 \min \{ L_x, L_y \},&\text{$\Lambda$ is a Type II lattice.}\\
\end{cases}
\end{eqnarray}
Thus, from Eq. (\ref{ngtc}), our two earlier stated main results follow:
{\it while for the more ``incommensurate'' Type I lattices, the
degeneracy will be  ``{\bf {topological}}" (i.e., given by $4^{g}$), for
Type II lattices, the degeneracy will be ``{\bf {holographic}}" (viz.,
the degeneracy will be exponential in the smallest of the edges along
the system boundaries)}. As discussed in Section \ref{sac}, the
additional constraints in Type II lattices are of the form of Eq.
(\ref{wl}). Expressed in terms of the four spin interaction terms
$A^z_{s}$ and $B^z_{p}$ of Eq. (\ref{H'}), a constraint of the form of
Eq. (\ref{wl}) states that there is a subset $\Gamma_{m} \subset
\Lambda$ for which $\prod _{s,p \in \Gamma_{m}} A^{z}_{s}B^{z}_{p}=1
$. An illustration of a constraint of such a type is provided, e.g. in
Fig. \ref{8-1.png}. Here, by virtue of the defining relations  of Eq.
(\ref{AB}), the product,
\begin{eqnarray}
A^{z}_{1}B^{z}_{1}A^{z}_{4}B^{z}_{4}=1.
\end{eqnarray}
Similarly, in panel a) of Fig. \ref{8-2.png}, colored arrows are drawn
along the diagonals. These colors code the constraints on the specific
$A_s^{z}$ and $B_p^{z}$ interaction terms. For example, along the {\it
green} arrows,
\begin{eqnarray}
\label{Green}
A^{z}_{1}B^{z}_{1}A^{z}_{4}B^{z}_{4}=1 \,\,\,\,\,\,\,\,\,\, \text{green (dashed)},
\end{eqnarray}  
and the constraints associated with the other  diagonals
\begin{eqnarray}
\label{Brown}
A^{z}_{2}B^{z}_{2}A^{z}_{3}B^{z}_{3}&=&1 \,\,\,\,\,\,\,\,\,\, \text{brown (dashed-dotted)},\nonumber\\
A^{z}_{2}B^{z}_{1}A^{z}_{3}B^{z}_{4}&=&1 \,\,\,\,\,\,\,\,\,\, \text{red (dashed-doubled-dotted)},\nonumber\\
A^{z}_{1}B^{z}_{2}A^{z}_{4}B^{z}_{3}&=&1 \,\,\,\,\,\,\,\,\,\, \text{black (dotted)} .
\end{eqnarray}
We provide another example in panel b) of Fig. \ref{8-2.png}. The
simplest visually appealing realization of Eq. (\ref{wl}) is that of the
subset $\Gamma_{m}$ being a trivial closed diagonal loop. Composites
(i.e., products) of independent constraints of the form of Eq.
(\ref{wl}) are, of course, also constraints. We aim to find the largest
number $(C^{\Lambda}_{g}-2)$ of such independent constraints. 
Non-trivial constraints formed by the product of bonds along real-space 
diagonal lines may appear. For example, in Fig. \ref{8-1.png}, the
product $A^{z}_{1}B^{z}_{1}A^{z}_{3}B^{z}_{2}=1$ is precisely such a
constraint. These constraints are more difficult to determine due to the
periodic boundary conditions. Generally, not all constraints are
independent of each other (e.g., multiplying any two constraints yield a
new constraint). The number of independent constraints, $C^{\Lambda}_g$
may be generally found by calculating the ``modular rank''  of the
linear equations formed by taking the logarithm of all constraints
found.   The qualified ``modular'' appears here as the $A^z_s$  and
$B^z_p$ eigenvalues  may only be $(\pm 1)$ and thus, correspondingly,
their phase is either $0$ or $\pi$. Many, yet generally, not all, of the
$C^{\Lambda}_g$ independent constrains are naturally associated with
products along the 45$^{\circ}$ lattice diagonals (as it appears on the
torus). Table \ref{table0} lists the numerically computed ground state degeneracies for
numerous lattices of genus $g=1$. All of these are concomitant with Eq.
(\ref{cdeg}). 

\begin{table}[htb]
\centering
\begin{tabular}{|c|c|c|c|c|c|l}
\cline{1-6}
Type & $L_x$ & $L_y$ & $E$ & $C_{g=1}^{\Lambda}$ & $n_{\sf{g.s.}}$ &\\ \cline{1-6}
\multicolumn{1}{ |c }{\multirow{3}{*}{I} } &
\multicolumn{1}{ |c| } 3 & 2 & 12 & 2 & 4      \\ \cline{2-6}
\multicolumn{1}{ |c  }{}                        &
\multicolumn{1}{ |c| } 5 & 2 & 20 & 2 & 4    \\ \cline{2-6}
\multicolumn{1}{ |c  }{}                        &
\multicolumn{1}{ |c| } 4 & 3 & 24 & 2 & 4     \\ \cline{2-6}
\multicolumn{1}{ |c  }{}                        &
\multicolumn{1}{ |c| } 5 & 3 & 30 & 2 & 4    \\ \cline{1-6}
\multicolumn{1}{ |c  }{\multirow{4}{*}{II} } &
\multicolumn{1}{ |c| } 2 & 2 & 8 & 4 & $4\times 2^2$   \\ \cline{2-6}
\multicolumn{1}{ |c  }{}                        &
\multicolumn{1}{ |c| } 4 & 2 & 16 & 4 & $4\times 2^2$    \\ \cline{2-6}
\multicolumn{1}{ |c  }{}                        &
\multicolumn{1}{ |c| } 6 & 2 & 24 & 4 & $4\times 2^2$    \\ \cline{2-6}
\multicolumn{1}{ |c  }{}                        &
\multicolumn{1}{ |c| } 3 & 3 & 18 & 6 & $4\times 2^4$  \\ \cline{2-6}
\multicolumn{1}{ |c  }{}                        &
\multicolumn{1}{ |c| } 4 & 4 & 32 & 8 & $4\times 2^6$  \\ \cline{1-6}
\end{tabular}
\caption{Computed ground state degeneracy ($n_{\sf{g.s.}}$) for  the classical
Toric Code  for different lattice sizes with genus one. Type I
corresponds to the case  $L_x \neq L_y$ where at least one of them is
odd. We put any other possibility under Type II which in general covers
the case $L_x \neq L_y$ where both $L_x$ and $L_y$ are even plus all
cases with $L_x=L_y$. In this table, $C_{g=1}^{\Lambda}$ denotes the
number of independent constraints (see text).}
\label{table0}
\end{table}

\subsection{Construction of ground states}
 
Given the symmetry operators of Eqs. (\ref{t+-}) and (\ref{tl}), we may
rather readily write down all ground states of the system. Denote the 
ferromagnetic ground state (i.e., one with all spins up ($|\!  \uparrow
\rangle_{( i j )}$) on all edges $( i j )$) by 
 \begin{eqnarray}
 \label{eF}
 |{\bf F} \rangle \equiv  \prod _{( ij )} | \uparrow\rangle_{( i j )};
 \end{eqnarray}
then, the four ground states of Type I lattices are
 \begin{eqnarray}
 \label{gn+}
 |G_{n_{+}, n_{-} }\rangle = (T^x_{+})^{n_{+}} (T^x_{-})^{n_{-}} | {\bf F} \rangle,
 \end{eqnarray}
where $n_{\pm} = 0,1$. Clearly, since $(T^x_{\pm})^{2} = 1$, only the
parity of the integers $n_{\pm}$ is important.  As (i) $[T^x_{\pm},
H]=0$ and (ii) the ferromagnetic state $|{\bf F} \rangle$ minimizes the
energy in  Eq. (\ref{H'}), it follows that all four binary strings
$(n_{+},n_{-}) = (0,0), (0,1), (1,0), (1,1)$ in Eq. (\ref{gn+}) lead to
$n_{\sf{g.s.}}=2^2 =4$  ground states.  The situation for Type II
lattices is a trivial extension of the above. That is, if there are
$(C_{g=1}^{\Lambda} - 2)$ additional independent symmetries $T^x_{m =
1}, T^x_{m =2}, \cdots, T^x_{m=  (C_{g=1}^{\Lambda} - 2)}$ of the
form of Eq. (\ref{tl}) then, with the convention of Eq. (\ref{eF}),  the
ground states will be  of the form
\begin{eqnarray}
\label{ggeneral}
&& |G_{n_{+} ,n_{-}, n_{1}, n_{2}, \cdots, n_{C_{g=1}^{\Lambda} - 2} }
\rangle = 
(T^x_{+})^{n_{+}} (T^x_{-})^{n_{-}} (T^x_{1})^{n_{1}}  \nonumber \\ 
&&\hspace*{2cm} \times (T^x_{2})^{n_{2}} \cdots  (T^x_{C_{g=1}^{\Lambda}-2})^{n_{
C_{g=1}^{\Lambda}-2}}  | {\bf F} \rangle,
\end{eqnarray}
with $2^{C_{g=1}^{\Lambda}}$ binary strings $(n_{+},n_{-},n_{1}, n_{2},
\cdots, n_{C_{g=1}^{\Lambda}-2})$, where $n_{m}=0,1$. These strings
span all possible $n_{\sf{g.s.}}= 2^{C_{g=1}^{\Lambda}}$ orthogonal
ground states.

Given the set of all orthonormal ground states $\{|g_{\alpha} \rangle
\}_{\alpha=1}^{n_{\sf{g.s.}}}$, it is possible to find quasi-local
operators ${\cal V}$ composed of $\sigma^z_{ij}$  ``operators'' on a
small number of edges such that 
 \begin{eqnarray}
 \label{TQO'}
 \langle g_{\alpha}|{\cal V}|g_{\alpha} \rangle = v_{\alpha} 
 \end{eqnarray}
 assumes different values  $v_{\alpha}$ in, at least, two different 
 ground states. 
 Equation (\ref{TQO'}) highlights that the expectation value of ${\cal
 V}$ is not state independent. In other words, Eq. (\ref{TQO})
 \cite{symmetry1, symmetry2,fragility} is violated. Thus, our classical
 system is, rather trivially, not topologically ordered .

\begin{figure}[htb]
\centering
\includegraphics[width=0.91 \columnwidth]{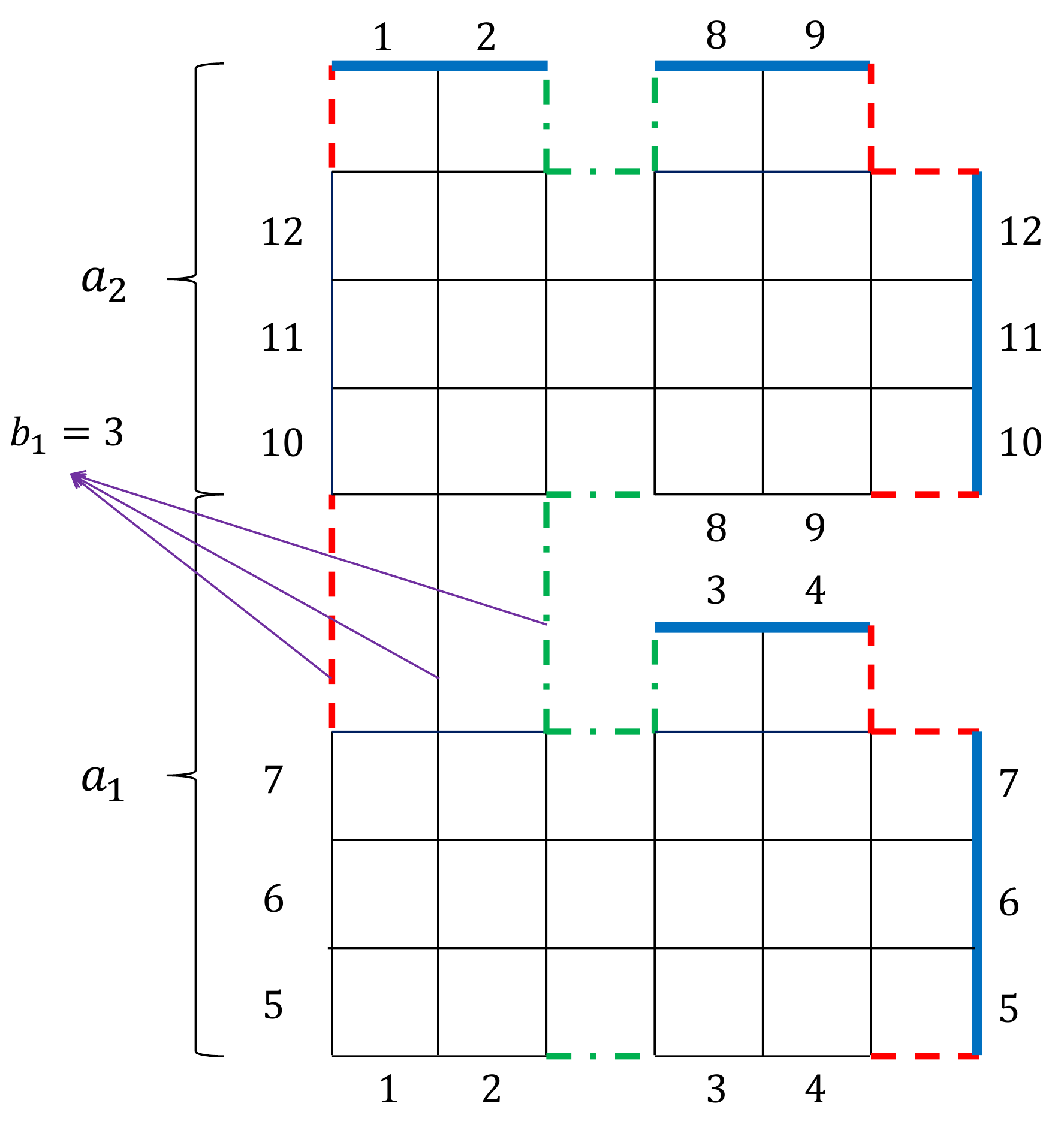}
\caption{A genus two ($g=2$) lattice. Identical bonds are labeled
by the same number (as a result of periodic boundary conditions). Thick solid
(blue) lines represent the boundary. The two plaquettes with $8$ bonds are
shown by dashed (red) and dashed-dotted (green) lines.}
\label{g=2.fig}
\end{figure}

\begin{figure}[htb]
\centering
\includegraphics[width=0.91 \columnwidth]{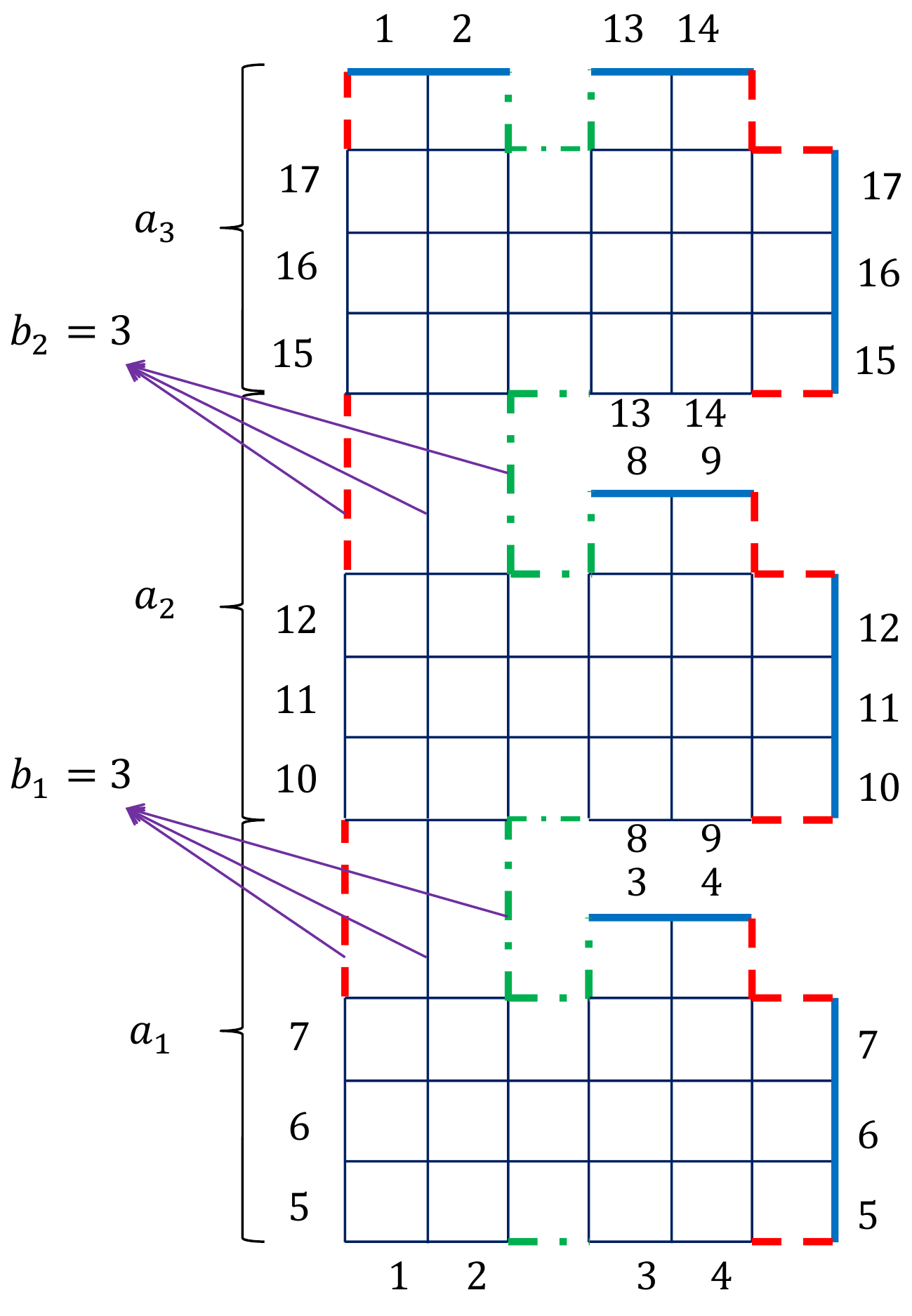}
\caption{A genus three ($g=3$) lattice. Identical bonds are labeled
by the same number (as a result of periodic boundary conditions). Thick solid
(blue) lines represent the boundary. The two plaquettes with $12$ bonds are
shown by dashed (red) and dashed-dotted (green) lines.}
\label{g=3.fig}
\end{figure}

\subsection{Ground state degeneracy on $g>1$ surfaces}
\label{construct_genus}

 Having understood the case of the simple torus ($g=1$), we will now
study lattices on surfaces $\Sigma$ of genus $g \ge 2$. We first explain
how to construct a finite size lattice of genus $g$
\cite{Ising-genus>1}.  Such lattices on genus $g$ ($g\geq 2$) surfaces
may be formed by {\it ``stitching together'' } $g$ simple parts $a_{j}$,
$j=1,\cdots,g$,  each of which largely looks like that of a simple
torus  (i.e., each region $a_j$ represents a set of vertices, edges and
faces of Type I  or II in the notation of Eq. (\ref{type12})),  via
$(g-1)$ ``bridges'' $\{b_{j}\}^{g-1}_{j=1}$. In Figs. \ref{g=2.fig}, and
\ref{g=3.fig}, the integer  number $b_j$ denotes the number of edges 
that regions $a_j$ and  $a_{j+1}$ share.

 To lucidly illustrate the basic construct, we start first with a $g=2$
lattice.  In Fig. \ref{g=2.fig}, identical edges are labeled by the same
number as a consequence of the periodic boundary conditions. Here, there
are $E=96$ edges, $V= 48$ vertices, and $F=46$ plaquettes.  As in the
case of the simple torus ($g=1$), the typical vertices are endpoints of
four edges. Similarly, in Fig. \ref{g=2.fig},  all plaquettes (with the
exception of two) are comprised of four edges as in the situation of the
simple torus. The exceptional cases are colored green (dashed-dotted)
and red (dashed). As seen in the figure, the lattice  may be splintered
into two regions (labeled by $a_1$ and $a_2$)  where one end of some of
the bonds belonging to $a_1$ are connected to $a_2$ as shown and labeled
in the picture under $b_1$. Each of the regions $a_1$ and $a_2$ looks,
by itself, like a square lattice on a genus $g=1$ surface. Generally,
the regions $a_1$ and $a_2$ may be composed of a different number of
edges. Employing the taxonomy of Eq. (\ref{type12}), we may classify
these regions $\{a_{j}\}_{j=1}^g$ to be of  either Type I or II. We
remark that the number of edges  $b_1$ must be always at least one less
than the minimum of the number of bonds of $a_1$ and $a_2$ along the
horizontal ($x$) axis. This algorithm trivially generalizes to higher
genus number.  The cartoon of Fig. \ref{g=3.fig} represents a lattice
with $g=3$.

 A synopsis of our numerical results  for the ground state degeneracy
for surfaces of genus $2 \le g \le 5$ appears in Table \ref{table1}. 
The ground state degeneracy depends on the type of each $a_j$ and the
number of bonds of each $b_j$. When all fragments $\{a_j\}$ are of Type
I and are inter-connected by only {\it single} common edges, the
degeneracy attains will its minimal possible value (Eq. (\ref{ngtc})) of
$4^{g}$ . 

\begin{table}[htb]
\centering
\begin{tabular}{|c|c|c|c|c|c|c|c|c|c|c|c|c|c}
\cline{1-13}
$g$ & $E$ & $n_{\sf{g.s.}}$ & Type &$a_1$ & $b_1$ & $a_2$ & $b_2$ & $a_3$ & $b_3$ & $a_4$ & $b_4$ & $a_5$ & 
 \\ \cline{1-13}
\multicolumn{1}{ |c }{\multirow{17}{*}{2} } &
\multicolumn{1}{ |c| } {8} & $4^g$ & 2 I& 2$\times$1& 1 & 2$\times$1& & & & & &  \\ \cline{2-7}
\multicolumn{1}{ |c  }{}                        &
\multicolumn{1}{ |c| } {10} & $4^g$ & 2 I& 3$\times$1& 1 & 2$\times$1& & & & & &    \\ \cline{2-7}
\multicolumn{1}{ |c  }{}                        &
\multicolumn{1}{ |c| } {12} & $4^g$ & 2 I& 3$\times$1& 1 & 3$\times$1& & & & & &    \\ \cline{2-7}
\multicolumn{1}{ |c  }{}                        &
\multicolumn{1}{ |c| } {16} & $4^g$ & 2 I& 3$\times$2& 1 & 2$\times$1& & & & & &    \\ \cline{2-7}
\multicolumn{1}{ |c  }{}                        &
\multicolumn{1}{ |c| } {18} & $4^g$ & 2 I& 3$\times$2& 1 & 3$\times$1& & & & & &   \\ \cline{2-7}
\multicolumn{1}{ |c  }{}                        &
\multicolumn{1}{ |c| } {24} & $4^g$ & 2 I& 3$\times$2& 1 & 3$\times$2& & & & & &    \\ \cline{2-7}
\multicolumn{1}{ |c  }{}                        &
\multicolumn{1}{ |c| } {24} & $4^g$ & 2 I& 5$\times$2& 1 & 2$\times$1& & & & & &    \\ \cline{2-7}
\multicolumn{1}{ |c  }{}                        &
\multicolumn{1}{ |c| } {12} &  $4^g\times2$ & 2 I& 3$\times$1& 2 & 3$\times$1& & & & &  & \\ \cline{2-7}
\multicolumn{1}{ |c  }{}                        &
\multicolumn{1}{ |c| } {12} &$4^g\times2$ & II+I& 2$\times$2& 1 & 2$\times$1& & & & & &    \\ \cline{2-7}
\multicolumn{1}{ |c  }{}                        &
\multicolumn{1}{ |c| } {14} & $4^g\times2$ & II+I& 2$\times$2& 1 & 3$\times$1& & & & & &    \\ \cline{2-7}
\multicolumn{1}{ |c  }{}                        &
\multicolumn{1}{ |c| } {20} & $4^g\times2$ & I+II& 3$\times$2& 1 & 2$\times$2& & & & & &    \\ \cline{2-7}
\multicolumn{1}{ |c  }{}                        &
\multicolumn{1}{ |c| } {20} & $4^g\times2$ & II+I& 4$\times$2& 1 & 2$\times$1& & & & & &    \\ \cline{2-7}
\multicolumn{1}{ |c  }{}                        &
\multicolumn{1}{ |c| } {22} & $4^g\times2$ & II+I& 4$\times$2& 1 & 3$\times$1& & & & & &    \\ \cline{2-7}
\multicolumn{1}{ |c  }{}                        &
\multicolumn{1}{ |c| } {24} & $4^g\times2$ & 2 I& 3$\times$2& 2 & 3$\times$2& & & & & &  \\ \cline{2-7}
\multicolumn{1}{ |c  }{}                        &
\multicolumn{1}{ |c| } {24} & $4^g\times2$ & II+I& 3$\times$3& 2 & 3$\times$1& & & & & &  \\ \cline{2-7}
\multicolumn{1}{ |c  }{}                        &
\multicolumn{1}{ |c| } {16} & $4^g\times2^3$ & 2 II& 2$\times$2& 1 & 2$\times$2& & & & & &   \\ \cline{2-7}
\multicolumn{1}{ |c  }{}                        &
\multicolumn{1}{ |c| } {24} & $4^g\times2^3$ & II+I& 3$\times$3& 1 & 3$\times$1& & & & & &   \\ \cline{2-7}
\multicolumn{1}{ |c  }{}                        &
\multicolumn{1}{ |c| } {24} & $4^g\times2^3$ & 2 II& 4$\times$2 & 1 & 2$\times$2& & & & & &    \\ \cline{1-13}
\multicolumn{1}{ |c }{\multirow{17}{*}{3} } &
\multicolumn{1}{ |c| } {12} & $4^g$ & 3 I& 2$\times$1& 1 & 2$\times$1& 1 & 2$\times$1& & & &   \\ \cline{2-9}
\multicolumn{1}{ |c  }{}                        &
\multicolumn{1}{ |c| } {14} & $4^g$ & 3 I& 3$\times$1& 1 & 2$\times$1& 1 & 2$\times$1 & & & &   \\ \cline{2-9}
\multicolumn{1}{ |c  }{}                        &
\multicolumn{1}{ |c| } {16} & $4^g$ & 3 I& 3$\times$1& 1 & 3$\times$1& 1 & 2$\times$1 & & & &  \\ \cline{2-9}
\multicolumn{1}{ |c  }{}                        &
\multicolumn{1}{ |c| } {16} &$4^g$ & 3 I& 3$\times$1& 2 & 3$\times$1& 1 & 2$\times$1 & & & &  \\ \cline{2-9}
\multicolumn{1}{ |c  }{}                        &
\multicolumn{1}{ |c| } {18} &$4^g$ & 3 I& 3$\times$1& 1 & 3$\times$1& 1 & 3$\times$1 & & & &   \\ \cline{2-9}
\multicolumn{1}{ |c  }{}                        &
\multicolumn{1}{ |c| } {18} & $4^g$ & 3 I& 3$\times$1& 2 & 3$\times$1& 1 & 3$\times$1 & & & &   \\ \cline{2-9}
\multicolumn{1}{ |c  }{}                        &
\multicolumn{1}{ |c| } {18} & $4^g$ & 3 I& 3$\times$1& 1 & 3$\times$1& 2 & 3$\times$1 & & & &   \\ \cline{2-9}
\multicolumn{1}{ |c  }{}                        &
\multicolumn{1}{ |c| } {18} & $4^g$ & 3 I& 3$\times$1& 2 & 3$\times$1& 2 & 3$\times$1 & & & &  \\ \cline{2-9}
\multicolumn{1}{ |c  }{}                        &
\multicolumn{1}{ |c| } {20} & $4^g$ & 3 I& 3$\times$2& 1 & 2$\times$1& 1 & 2$\times$1 & & & &   \\ \cline{2-9}
\multicolumn{1}{ |c  }{}                        &
\multicolumn{1}{ |c| } {24} & $4^g$ & 3 I& 3$\times$2& 1 & 3$\times$1& 1 & 3$\times$1 & & & &   \\ \cline{2-9}
\multicolumn{1}{ |c  }{}                        &
\multicolumn{1}{ |c| } {24} & $4^g$ & 3 I& 3$\times$2& 2 & 3$\times$1& 2 & 3$\times$1 & & & &   \\ \cline{2-9}
\multicolumn{1}{ |c  }{}                        &
\multicolumn{1}{ |c| } {16} & $4^g\times2$ & 2 I+II& 2$\times$1& 1 & 2$\times$1& 1 & 2$\times$2 & & & &   \\ \cline{2-9}
\multicolumn{1}{ |c  }{}                        &
\multicolumn{1}{ |c| } {18} & $4^g\times2$ & 2 I+II& 3$\times$1& 1 & 2$\times$1& 1 & 2$\times$2 & & & &   \\ \cline{2-9}
\multicolumn{1}{ |c  }{}                        &
\multicolumn{1}{ |c| } {20} & $4^g\times2$ & 2 I+II& 3$\times$1& 1 & 3$\times$1& 1 & 2$\times$2 & & & &   \\ \cline{2-9}
\multicolumn{1}{ |c  }{}                        &
\multicolumn{1}{ |c| } {20} &$4^g\times2^2$ & 2 II+I& 2$\times$2& 1 & 2$\times$2& 1 & 2$\times$1 & & & &   
\\ \cline{2-9}
\multicolumn{1}{ |c  }{}                        &
\multicolumn{1}{ |c| } {22} & $4^g\times2^2$ & 2 II+I& 2$\times$2& 1 & 2$\times$2& 1 & 3$\times$1 & & & &   
\\ \cline{2-9}
\multicolumn{1}{ |c  }{}                        &
\multicolumn{1}{ |c| } {24} & $4^g\times2^4$ &3 II& 2$\times$2& 1 & 2$\times$2& 1 & 2$\times$2 & & & &  
\\ \cline{1-13}
\multicolumn{1}{ |c }{\multirow{4}{*}{4} } &
\multicolumn{1}{ |c| } {16} & $4^g$ & 4 I&  2$\times$1& 1 & 2$\times$1& 1 & 2$\times$1 & 1 & 2$\times$1 & &  
\\ \cline{2-11}
\multicolumn{1}{ |c  }{}                        &
\multicolumn{1}{ |c| } {18} & $4^g$ & 4 I&  2$\times$1& 1 & 2$\times$1& 1 & 2$\times$1 & 1 & 3$\times$1 & & 
\\ \cline{2-11}
\multicolumn{1}{ |c  }{}                        &
\multicolumn{1}{ |c| } {24} & $4^g$ & 4 I&  3$\times$2& 1 & 2$\times$1& 1 & 2$\times$1 & 1 & 2$\times$1 & & 
\\ \cline{2-11}
\multicolumn{1}{ |c  }{}                        &
\multicolumn{1}{ |c| } {20} & $4^g\times2$ & II+3 I & 2$\times$2& 1 & 2$\times$1& 1 & 2$\times$1 & 1 & 2$\times$1 
& &  \\ \cline{1-13}
\multicolumn{1}{ |c }{\multirow{2}{*}{5} } &
\multicolumn{1}{ |c| } {20} & $4^g$ & 5 I& 2$\times$1& 1 & 2$\times$1& 1 & 2$\times$1 & 1 & 2$\times$1 &1 & 
2$\times$1   \\ \cline{2-13}
\multicolumn{1}{ |c  }{}                        &
\multicolumn{1}{ |c| } {24} & $4^g\times2$ & II + 4 I & 2$\times$2& 1 & 2$\times$1& 1 & 2$\times$1 & 1 & 
2$\times$1 &1 & 2$\times$1   \\ \cline{1-13}
\end{tabular}
\caption{Computed ground state degeneracy ($n_{\sf{g.s.}}$) for square lattices  with $g>1$. 
The $g$ denotes ``genus'' (see text).}
\label{table1}
\end{table}

If, in Eq. \eqref{H'}, we set $J$ to zero, we will obtain the
Hamiltonian  of the Ising gauge model. As this theory does not have a
star term, this Hamiltonian involves more symmetries and, therefore, one
expects the ground state subspace to have a larger degeneracy.  We
numerically verified it to be $n^{\sf gauge}_{\sf g.s.}= 
4^{g}\times 2^{\sf N_{site}-1}$-fold degenerate (${\sf N_{site}}= E/2$) \cite{local-gauge-Ising}.
 
\begin{figure}[b]
\centering
\includegraphics[width=0.91 \columnwidth]{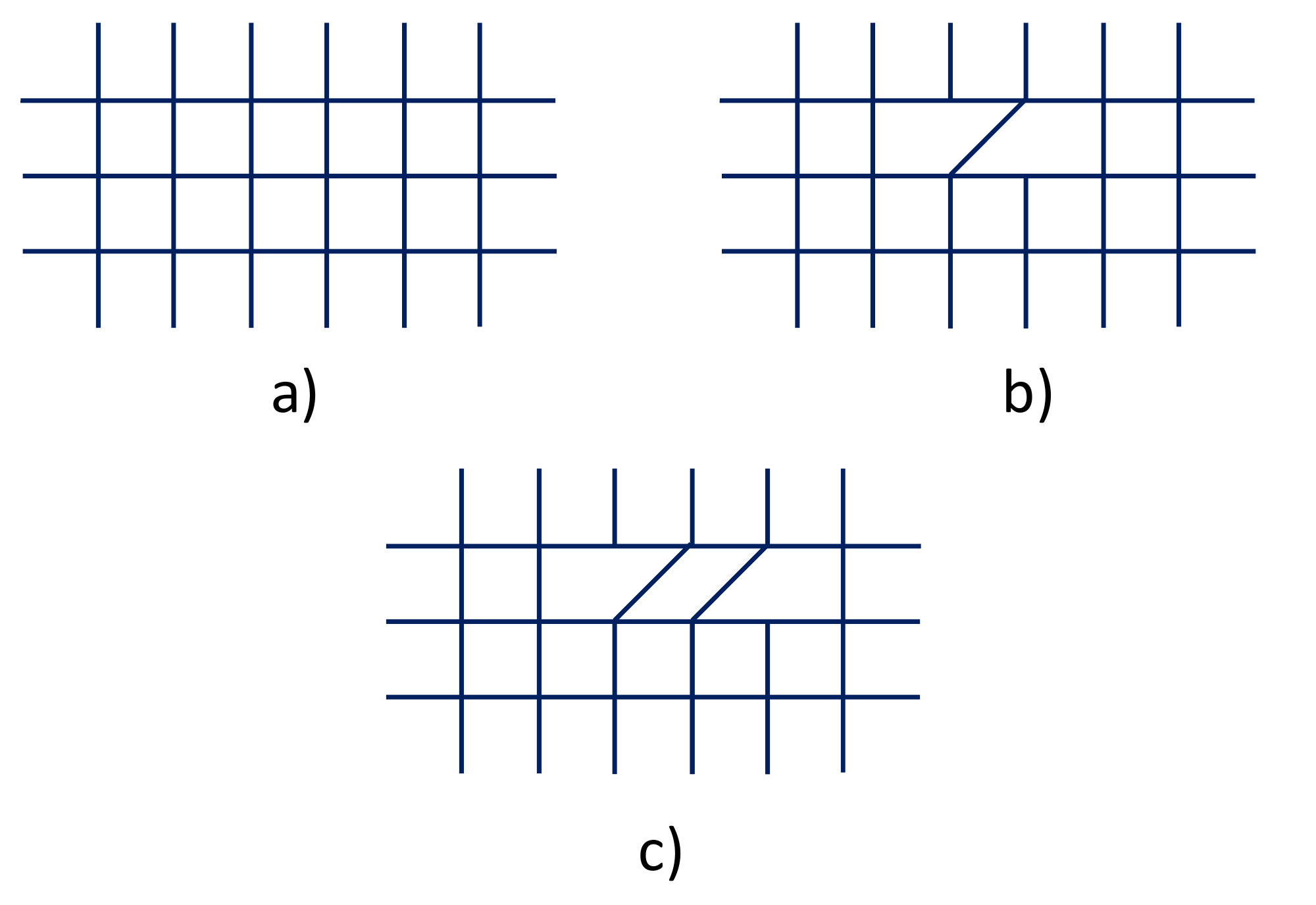}
\caption{Sketch of a part of a square lattice a) with two types of
defects b) and c). The defective lattices in b) and c) have one bond
less than in a).}
\label{Defect.png}
\end{figure}

\begin{table}[htb]
\centering
\begin{tabular}{|c|c|c|c|c|c|c|c|c|c|c|c|}
\cline{1-11}
$g$ & $E$ & $n_{\sf{g.s.}}$ & Type &$a_1$ & $b_1$ & $a_2$ & $b_2$ & $a_3$ & $b_3$ & $a_4$  \\ \cline{1-11}
\multicolumn{1}{ |c }{\multirow{8}{*}{1} } &
\multicolumn{1}{ |c| } {11} & $4^g$ & I & 3$\times$2 $\star$& & & & & &     \\ \cline{2-5}
\multicolumn{1}{ |c  }{}                        &
\multicolumn{1}{ |c| } {15} & $4^g$ & II & 4$\times$2 $\star$& & & & & &     \\ \cline{2-5}
\multicolumn{1}{ |c  }{}                        &
\multicolumn{1}{ |c| } {19} & $4^g$ & I & 5$\times$2 $\star$& & & & & &    \\ \cline{2-5}
\multicolumn{1}{ |c  }{}                        &
\multicolumn{1}{ |c| } {23} & $4^g$ & I & 6$\times$2 $\star$ & & & & & &     \\ \cline{2-5}
\multicolumn{1}{ |c  }{}                        &
\multicolumn{1}{ |c| } {23} & $4^g$ & I & 4$\times$3 $\star$& & & & & &   \\ \cline{2-5}
\multicolumn{1}{ |c  }{}                        &
\multicolumn{1}{ |c| } {16} & $4^g\times2$ & II & 3$\times$3 $2\star$& & & & & &     \\ \cline{2-5}
\multicolumn{1}{ |c  }{}                        &
\multicolumn{1}{ |c| } {17} & $4^g\times2$ & II & 3$\times$3 $\star$& & & & & &     \\ \cline{2-5}
\multicolumn{1}{ |c  }{}                        &
\multicolumn{1}{ |c| } {19} & $4^g\times2$ & I & 5$\times$2 $\star\star$& & & & & &    \\ \cline{2-5}
\multicolumn{1}{ |c  }{}                        &
\multicolumn{1}{ |c| } {22} & $4^g\times2$ & I & 6$\times$2 $2\star$& & & & & &    \\ \cline{1-11}
\multicolumn{1}{ |c }{\multirow{9}{*}{2} } &
\multicolumn{1}{ |c| } {15} & $4^g$ & 2 I& 3$\times$2 $\star$& 1 & 2$\times$1& & & &     \\ \cline{2-7}
\multicolumn{1}{ |c  }{}                        &
\multicolumn{1}{ |c| } {17} & $4^g$ & 2 I& 3$\times$2 $\star$& 1 & 3$\times$1& & & &   \\ \cline{2-7}
\multicolumn{1}{ |c  }{}                        &
\multicolumn{1}{ |c| } {21} & $4^g$ & 2 I& 4$\times$2 $\star$& 1 & 3$\times$1& & & &   \\ \cline{2-7}
\multicolumn{1}{ |c  }{}                        &
\multicolumn{1}{ |c| } {22} & $4^g$ & 2 I& 3$\times$2 $\star$& 1 & 3$\times$2 $\star$& & & & \\ \cline{2-7}
\multicolumn{1}{ |c  }{}                        &
\multicolumn{1}{ |c| } {23} & $4^g$ & 2 I& 3$\times$2 $\star$& 1 & 3$\times$2& & & &  \\ \cline{2-7}
\multicolumn{1}{ |c  }{}                        &
\multicolumn{1}{ |c| } {23} & $4^g$ & 2 II& 4$\times$2 $\star$& 1 & 2$\times$2& & & &   \\ \cline{2-7}
\multicolumn{1}{ |c  }{}                        &
\multicolumn{1}{ |c| } {23} & $4^g$ & II+I& 3$\times$3 $\star$& 2 & 3$\times$1& & & &  \\ \cline{2-7}
\multicolumn{1}{ |c  }{}                        &
\multicolumn{1}{ |c| } {23} & $4^g$ & 2 I& 5$\times$2 $\star$& 1 & 2$\times$1& & & &  \\ \cline{2-7}
\multicolumn{1}{ |c  }{}                        &
\multicolumn{1}{ |c| } {23} & $4^g\times2$ & II+I& 3$\times$3 $\star$& 1 & 3$\times$1& & & &   \\ \cline{1-11}
\multicolumn{1}{ |c }{\multirow{3}{*}{3} } &
\multicolumn{1}{ |c| } {19} & $4^g$ & 3 I& 3$\times$2 $\star$& 1 & 2$\times$1& 1 & 2$\times$1 & &   \\ \cline{2-9}
\multicolumn{1}{ |c  }{}                        &
\multicolumn{1}{ |c| } {23} & $4^g$ & 3 I& 3$\times$2 $\star$& 1 & 3$\times$1& 1 & 3$\times$1 & & \\ \cline{2-9}
\multicolumn{1}{ |c  }{}                        &
\multicolumn{1}{ |c| } {23} & $4^g$ & 3 I& 3$\times$2 $\star$& 2 & 3$\times$1& 2 & 3$\times$1 & &  \\ \cline{1-11}
\multicolumn{1}{ |c }{\multirow{1}{*}{4} } &
\multicolumn{1}{ |c| } {23} & $4^g$ & 4 I&  3$\times$2 $\star$& 1 & 2$\times$1& 1 & 2$\times$1 & 1 & 2$\times$1 
\\ \cline{1-11}
\end{tabular}
\caption{Computed ground state degeneracy ($n_{\sf{g.s.}}$) of defective square
lattices. The $g$ denotes  ``genus''.  By ``$2\star$'' we mean there are
$2$ defects of type ``$\star$'' (see text).}
\label{table2}
\end{table}

\subsection{Lattice Defects}

When dislocations and/or any other lattice defects are present in the
classical Toric Code model, the degeneracy is, of course, still bounded
by the geometry independent result of $4^{g}$. On Type I
lattice (and their composites), the degeneracy is typically equal to
this bound yet it may go up upon the introduction of defects. Similarly,
in most cases introducing such lattice defects lowers the degeneracy of
the more commensurate Type II lattices (and their composites).

Table \ref{table2} provides the numerical results for such defective
lattices.  For example, in Fig. \ref{Defect.png} we see the original
lattice, panel a), along with two types of defects as in panel b) and
c).  These are obtained by replacing $3$ squares by $2$ adjacent or
separated pentagons as in panel b) and c), respectively. To avoid
confusion, we will use ``$\star$'' sign for the first case and
``$\star\star$'' for the second case. By putting a ``$\star$''
(``$\star\star$'') sign beside a $3\times2$ lattice, we mean it exhibits
a defect of type one (two). That is, represented as ``$3\times2 \,\,
\star$'' (``$3\times2 \,\, \star\star$'').

\section{Thermodynamics of the Classical Toric Code Model}
\label{CTC}

Previous sections largely focused on the ground states of the classical
Toric Code model. As our earlier considerations make clear, however, a
minimal topology (and general constraint) dependent degeneracy
${\cal{N}}_{\sf global} \equiv \min(n_{ \sf g.s.})$ appears for all
levels (see, e.g., Eq. (\ref{mindeg})). This ``global'' degeneracy must
manifest itself as a prefactor in the computation of the partition
function. That is, if the whole  spectrum has a global degeneracy
${\cal{N}}_{\sf global}$ then the canonical partition function may be
expressed as
\begin{eqnarray}
\label{ctczl}
{\cal Z} = {\cal{N}}_{\sf global} \sum_{\ell=0} n_{\ell} e^{-\beta E_{\ell}},
\end{eqnarray}
where ${\cal{N}}_{\sf global} \, n_{\ell} \ge {\cal{N}}_{\sf global}$ is the number of states
having total energy $E_{\ell}$.  In ``incommensurate'' lattices, when
no constraints $\{{\cal{C}}_{m}\}$ augment those of Eq. (\ref{ABC}),
we find that, similar to the partition function of the quantum Toric
Code model \cite{symmetry1,symmetry2,fragility},  the partition of the
classical Toric Code model is given by
\begin{eqnarray}
\label{ctcz}
{\cal Z}_{\sf inc.} =&& 4^{g-1}[(2 \cosh \beta J)^{V} + (2 \sinh \beta
J)^{V}] \nonumber \\ 
&& \times [ (2 \cosh \beta J')^{F} + (2 \sinh \beta J')^{F}].
\end{eqnarray}
The prefactor of $4^{g-1}$ embodies the increase in degeneracy by a
factor of four as $g$ is elevated in increments $g \to (g+1)$ beyond a
value of $g=1$. On the simple torus (i.e., when $g=1$), this partition
function (similar to the partition function of the quantum Toric Code
model \cite{symmetry1,symmetry2,fragility}) is that of two decoupled
Ising chains with one of these chains having $V$ spins and the other
composed of $F$ spins. As each such Ising chain has a two-fold
degeneracy, it thus follows that the degeneracy of the (more
``incommensurate'') Type I $g=1$ system is four-fold and that the
degeneracy of the classical Toric Code model on incommensurate lattices
on Riemann surfaces of genus $g$ is $4^{g}$ for all $g \ge 1$. The
latter value saturates the lower bound on the degeneracy of Eq.
(\ref{mindeg}). In Appendix \ref{Z}, we list the partition function for
several other more commensurate finite size lattice realizations.

\section{Classical Toric Clock Models and their Clock gauge theory limits}
\label{sec:clock}

In this section, we introduce and study a clock model ($\mathbb{Z}_{{\sf
d}_{Q}}$) extension of the classical Toric Code model. To that end, we
consider what occurs when each spin $S$ may assume ${\sf d}_Q > 2$
values.  Specifically, on every oriented ($i \rightarrow j$) edge (that
we will hereafter  label as $(ij)$), we set
\begin{eqnarray}
\label{sijk}
\sigma_{ij}&=& \exp\Big[{\rm i}\, \frac{2\pi}{{\sf d}_Q}  \alpha_{ij} \Big ],\,\,\,\,
(\alpha_{ij} = 0,1,\cdots,{\sf d}_Q-1), \\
\sigma_{ji}&=& \sigma^{*}_{ij}.
\end{eqnarray}
The last equality reflects that a change in the orientation (i.e., a
link in the  direction from $j \to i$ as opposed to $i \to j$) is
associated with complex conjugation. At each vertex ``$s$'', we define
$A_s$ as
\begin{eqnarray}
\label{NewA}
A_{s}&=&\frac{1}{2}
(\sigma_{si}^{\;}\sigma_{sj}\sigma_{sk}^{\;}\sigma_{sl} +
{\rm H.c.}) \nonumber\\
&=& \cos \Big(
\frac{2\pi}{{\sf d}_Q}(\alpha_{si}+\alpha_{sj}+\alpha_{sk}+\alpha_{sl}) \Big),
\end{eqnarray}
and for each plaquette $p$
\begin{eqnarray}
\label{NewB}
B_{p}=\cos \Big(
\frac{2\pi}{{\sf d}_Q}(\alpha_{ij}+\alpha_{jk}+\alpha_{kl}+\alpha_{li}) \Big),
\end{eqnarray}
composed of edges $(ij), (jk), (kl), (li)$, such that the loop $i
\rightarrow j \rightarrow k \rightarrow l$ is oriented 
counter-clockwise around about the plaquette center.   Table
\ref{table3} provides our numerical results for ground state degeneracy
($D^{0}_{{\sf d}_Q}$) for different size lattices of varying genus
numbers $g$. The ${\sf d}_Q=2$ case is that  investigated in the earlier
sections (i.e., that of the classical Toric Code model with Ising
variables $\sigma_{ij} = \pm 1$). 

It is readily observed that the {\it minimal} ground state degeneracy is
set by the genus number,
\begin{eqnarray}
\label{Min}
n^{\sf min}_{\sf g.s.} = \min\{ D^{0}_{{\sf d}_Q}\}= 
\begin{cases}
	{\sf d}_{Q}^{2g-1},&\text{odd ${\sf d}_Q$}, \\
	&\\
	2 {\sf d}_{Q}^{2g-1} ,&\text{even ${\sf d}_Q$}.\\
\end{cases}
\end{eqnarray}
We next introduce a simple framework that rationalizes Eq. (\ref{Min})
and enables us to furthermore derive the results of the previous
sections (i.e., the Ising case of ${\sf d}_{Q} =2$) in a unified way.
Furthermore, this approach will allow us to better understand not only
the degeneracies in the ground sector but also those of all higher
energy states. In the up and coming, we will study the Hamiltonian
\begin{eqnarray}
\label{CGEnergy}
H_{{\sf d}_Q} &=& -\sum_{s}A_s -\sum_{p}B_p \\
&=& -\sum_{s}\cos \Big( \ \frac{2\pi m_{s,{\sf d}_Q}}{{\sf d}_Q} \Big)
-\sum_{p}\cos  \Big( \frac{2\pi   m_{p,{\sf d}_Q}}{{\sf d}_Q} \Big). \nonumber
\end{eqnarray}
Here,
\begin{eqnarray}
\label{AlphaEq}
	\begin{cases}
		m_{s,{\sf d}_Q}=\alpha_{si}+\alpha_{sj}+\alpha_{sk}+\alpha_{sl},\\
		m_{p,{\sf d}_Q}=\alpha_{ij}+\alpha_{jk}+\alpha_{kl}+\alpha_{li}, \\
	\end{cases}
\end{eqnarray}
constitute a system of linear equations. A pair of fixed integers $m^{\ell}_{s,{\sf d}_Q}$ 
and $m^{\ell}_{p,{\sf d}_Q}$ defines an energy  $E_\ell$. There are 
$n^{\ell}_{{\sf d}_Q}$ such pairs. 

For each fixed pair $r$, $r=1,\cdots,n^{\ell}_{{\sf d}_Q}$, we may 
express these linear equations as
\begin{eqnarray}
\label{Matrix}
WX^{r} = Y^{r} ,
\end{eqnarray} 
where  $W$ is a rectangular ($(V+F)\times E$) matrix.
The matrix elements of $W$ are either $0$ or $\pm 1$. Generally, the
form of the matrix $W$ depends on both the size and type of  lattice.
The dimension of the vector $X^{r}$ is equal to the number ($E$) of
edges; $Y^{r}$ is a $(V+F)-$component vector. Specifically, following
Eq. (\ref{AlphaEq}), these two vectors are defined as: 
$X^{r}=\vec{\alpha}$, with components $\alpha_{ij}$, and 
$Y^r=m^{\ell}_{s,{\sf d}_Q}$, for its first $V$ components and 
$Y^r=m^{\ell}_{p,{\sf d}_Q}$, for the remaining $F$ components.

The number of linearly independent equations ($r_{{\sf d}_Q}$) is equal
to the rank of the matrix $W$. Typically, the rank
$r_{{\sf d}_Q}$ is less than the number of unknown $\alpha_{ij}$.
Therefore, we cannot determine all $\alpha_{ij}$ from Eq.
(\ref{Matrix}).  We should note that the rank of the matrix $W$ is
computed modularly, ``$\bmod~{\sf d}_Q$''. This latter modular rank is
of pertinence as the edge variables $\alpha_{ij}$ may only take on
particular modular values ($\alpha_{ij}=0,1,\cdots,{\sf d}_Q-1$).

\onecolumngrid
\begin{center}
\begin{table}[htb]
\centering
\begin{tabular}{|c|c|c|c|c|c|c|c|c|c|c|c|c|c|c|c|c|c|c|c|c|c|c|c|c}
\cline{1-24} 
$g$ & $E$ & Type &$a_1$ & $b_1$ & $a_2$ & $b_2$ & $a_3$ & $b_3$ & $a_4$ & $\sf N^{0}_{3}$& $\sf N^{0}_{4}$& $\sf N^{0}_{5}$& $\sf N^{0}_{6}$ & $\sf N^{0}_{7}$ & $\sf N^{0}_{8}$ & $\sf N^{0}_{9}$ & $\sf N^{0}_{10}$ & $\sf N^{0}_{11}$ & $\sf N^{0}_{12}$ & $\sf N^{0}_{13}$ & $\sf N^{0}_{14}$ & $\sf N^{0}_{15}$ & $\sf N^{0}_{16}$  \\ \cline{1-24}
\multicolumn{1}{ |c }{\multirow{6}{*}{1} } &
\multicolumn{1}{ |c| } {4} & I & 2$\times$1 & & & & & & & 1 & 2 & 1 & 1 & 1 & 2 & 1 & 1 & 1 & 2 & 1 & 1 & 1 & 2  \\ \cline{2-24} 
\multicolumn{1}{ |c  }{}                        &
\multicolumn{1}{ |c| } {6} & I & 3$\times$1 & & & & & & &3 & 1 & 1 & 3 & 1& 1 & 3 & 1 & 1& 3 & 1 & 1 & 3 & 1   \\ \cline{2-24}
\multicolumn{1}{ |c  }{}                        &
\multicolumn{1}{ |c| } {8} & I & 4$\times$1 & & & & & & & 1 & 2 & 1 & 1 & 1& 4 & 1 & 1 & 1& 2 & 1 & 1 & &   \\ \cline{2-22}
\multicolumn{1}{ |c  }{}                        &
\multicolumn{1}{ |c| } {8} & II & 2$\times$2 & & & & & & & $3^2$ & $4^2$ & $5^2$ & $6^2$ & $7^2$& $8^2$ & $9^2$ & $10^2$ & $11 ^2$ & $12^2$ & $13^2$ & $14^2$ & &   \\ \cline{2-22}
\multicolumn{1}{ |c  }{}                        &
\multicolumn{1}{ |c| } {12} & I & 3$\times$2 & & & & & & & 3 & 2 & 1 & & & & & & & & & & &   \\ \cline{2-13}
\multicolumn{1}{ |c  }{}                        &
\multicolumn{1}{ |c| } {16} & II & 4$\times$2 & & & & & & & $3^2$ & $2\times 4^2$ & & & & & & & & & & & &   \\ \cline{2-12}
\multicolumn{1}{ |c  }{}                        &
\multicolumn{1}{ |c| } {18} & II & 3$\times$3 & & & & & & & $3^4$ & & & & & & & & & & & & &   \\ \cline{1-24}
\multicolumn{1}{ |c }{\multirow{5}{*}{2} } &
\multicolumn{1}{ |c| } {8} & 2 I& 2$\times$1 & 1 & 2$\times$1& & & & & 1 & 2 & 1 & 1 & 1& 2 & 1 & 1 & 1 & 2 & 1 & 1 & &  \\ \cline{2-22}
\multicolumn{1}{ |c  }{}                        &
\multicolumn{1}{ |c| } {12} & 2 I& 3$\times$1 & 1 & 3$\times$1& & & & & 3 & 1 & 1 & & & & & & & & & & &  \\ \cline{2-13}
\multicolumn{1}{ |c  }{}                        &
\multicolumn{1}{ |c| } {12} & 2 I& 3$\times$1 & 2 & 3$\times$1& & & & & 3 & 2 & 1 & & & & & & & & & & &  \\ \cline{2-13}
\multicolumn{1}{ |c  }{}                        &
\multicolumn{1}{ |c| } {12} & II+I& 2$\times$2 & 1 & 2$\times$1& & & & & 1 & 4 & 1 & & & & & & & & & & &  \\ \cline{2-13}
\multicolumn{1}{ |c  }{}                        &
\multicolumn{1}{ |c| } {16} & 2 II& 2$\times$2 & 1 & 2$\times$2& & & & & $3^2$ & $2\times 4^2$  & & & & & & & & & & & &  \\ \cline{2-12}
\multicolumn{1}{ |c  }{}                        &
\multicolumn{1}{ |c| } {18} & 2 I& 3$\times$2 & 1 & 3$\times$1& & & & & 3 & & & & & & & & & & & & &  \\ \cline{1-24}
\multicolumn{1}{ |c }{\multirow{3}{*}{3} } &
\multicolumn{1}{ |c| } {12} & 3 I& 2$\times$1 & 1 & 2$\times$1& 1 & 2$\times$1& & & 1 & 2 & 1 & & & & & & & & & & &    \\ \cline{2-13}
\multicolumn{1}{ |c  }{}                        &
\multicolumn{1}{ |c| } {16} & 3 I& 3$\times$1 & 1 & 3$\times$1& 1 & 2$\times$1& & & 1 & 1 & & & & & & & & & & & &    \\ \cline{2-12}
\multicolumn{1}{ |c  }{}                        &
\multicolumn{1}{ |c| } {16} & 2 I+II& 2$\times$1 & 1 & 2$\times$1& 1 & 2$\times$2& & & 1 & 4 & & & & & & & & & & & &    \\ \cline{2-12}
\multicolumn{1}{ |c  }{}                        &
\multicolumn{1}{ |c| } {18} & 2 I+II& 3$\times$1 & 1 & 2$\times$1& 1 & 2$\times$2& & & 1 & & & & & & & & & & & & &    \\ \cline{1-24}
\multicolumn{1}{ |c }{\multirow{2}{*}{4} } &
\multicolumn{1}{ |c| } {16} & 4 I&  2$\times$1 & 1 & 2$\times$1& 1 & 2$\times$1 & 1 & 2$\times$1 & 1 & 2 & & & & & & & & & & & &\\ \cline{2-12}
\multicolumn{1}{ |c  }{}                        &
\multicolumn{1}{ |c| } {18} & 4 I&  2$\times$1 & 1 & 2$\times$1& 1 & 2$\times$1 & 1 & 3$\times$1 & 1 & & & & & & & & & & & & &\\ \cline{1-24}
\end{tabular}
\caption{Computed departure from the minimal ground state degeneracy, ${\sf N}^{0}_{\sf M}$ = 
${D^{0}_{\sf M}}/{n^{\sf min}_{\sf g.s.}}$, where
$D^{0}_{\sf M}$ denotes the ground state degeneracy for ${\sf d}_Q={\sf M}$, and
$n^{\sf min}_{\sf g.s.}$ is equal to ${\sf d}_{Q}^{2g-1}$ ($2 {\sf
d}_{Q}^{2g-1}$) for odd (even) ${\sf d}_{Q}$.}
\label{table3}
\end{table}
\end{center}
\twocolumngrid

Our objective is to calculate the degeneracy $D^{\ell}_{{\sf d}_Q}$ of each
energy level $\ell$ (or sector of states that share the same energy of Eq.
(\ref{CGEnergy})). Equation (\ref{Matrix}) imposes $r_{{\sf d}_Q}$
constraints on the ${\sf d}_{Q}$ possible values of $\alpha_{ij}$. Thus, for
each set of integers $m^{\ell}_{s,{\sf d}_Q}$  and $m^{\ell}_{p,{\sf d}_Q}$,
the degeneracy is equal to ${\sf d}_Q^{E-r_{{\sf d}_Q}}$. As there are
$n^{\ell}_{{\sf d}_Q}$ such sets of integers (see Eq. (\ref{Matrix})), the
degeneracy of each level $\ell$ is
\begin{eqnarray}
\label{dld}
D^{\ell}_{{\sf d}_Q}= n^{\ell}_{{\sf d}_Q} {\sf d}_Q^{E-r_{{\sf d}_Q}}.
\end{eqnarray}
We may recast Eq. (\ref{dld}) to highlight the effect of topology and
invoke the Euler relation (Eqs. (\ref{c22}) and (\ref{Euler})) to write the
degeneracy as
\begin{eqnarray}
\label{Deg-Gen}
D^{\ell}_{{\sf d}_Q}= n^{\ell}_{{\sf d}_Q} {\sf d}_Q^{2(g-1)+C^{\Lambda}_{g}},
\end{eqnarray}  
where we define
\begin{eqnarray}
\label{CLambda1}
C^{\Lambda}_{g} \equiv V+F-r_{{\sf d}_Q}.
\end{eqnarray}
The modular rank of the matrix $W$ lies in the interval $1\leq r_{{\sf d}_Q} <
V+F$. It thus follows that
\begin{eqnarray}
\label{CLambda2}
1\leq C^{\Lambda}_{g} \leq V+F-1.
\end{eqnarray}
From Eqs. (\ref{Deg-Gen}) and (\ref{CLambda2}), it is readily seen that
\begin{eqnarray}
\label{mindg}
D^{\ell}_{{\sf d}_Q} \geq {\sf d}_Q^{2g-1}.
\end{eqnarray}
The degeneracy of Eq. (\ref{mindg}) (stemming from the spectral
redundancy of each level $\ell$ seen in Eq. (\ref{Deg-Gen})) is consistent
with an effective composite symmetry 
\begin{eqnarray}
\label{zdq}
G=  \mathbb{Z}_{{\sf d}_Q} \otimes\mathbb{Z}_{{\sf d}_Q} \otimes \cdots \otimes 
\mathbb{Z}_{{\sf d}_Q},
\end{eqnarray}
i.e., the product of $(2g-1)$ symmetries of the $\mathbb{Z}_{{\sf d}_Q}$
type.   That is, if each element of such a $\mathbb{Z}_{{\sf d}_Q}$
symmetry gave rise  to a ${\sf d}_Q$-fold degeneracy then the result of
Eq. (\ref{Deg-Gen}) will naturally follow. 

The non-local symmetry of Eq.
(\ref{zdq}) compound the standard local symmetries that appear in the
gauge theory limit of Eq. (\ref{CGEnergy}) in which the $A_{s}$  terms
are absent, i.e., $H_{{\sf d}_Q}=
-\sum_{p}B_p$. The latter  gauge theory enjoys the local symmetries
\begin{eqnarray}
 \theta_{ij} \to \theta_{ij} + \phi_{i} - \phi_{j},
 \label{phaseijclock}
\end{eqnarray}
with, at any lattice vertex (site) $i$, the angle $\phi_{i}$ being an arbitrary
integer multiple of  ${2\pi}/{{\sf d}_Q}$. 
In this case, we find that the ground state degeneracy 
($D^{\sf gauge, 0}_{\sf d_Q}$) is purely topological (i.e., not holographic),
\begin{eqnarray}
\label{Min_gauge}
D^{\sf gauge, 0}_{{\sf d}_Q} = n^{\sf gauge,0}_{{\sf d}_Q} {\sf d}_Q^{2(g-1)+ \frac{E}{2}},
\end{eqnarray}
where,
\begin{eqnarray}
\label{Min_gauge_Con}
\begin{cases}
	1 \leq n^{\sf gauge,0}_{{\sf d}_Q} \leq {\sf d}_{Q},&\text{odd ${\sf d}_Q$}, \\
	&\\
	2 \leq n^{\sf gauge,0}_{{\sf d}_Q} \leq {\sf d}_{Q} ,&\text{even ${\sf d}_Q$}.\\
\end{cases}
\end{eqnarray}
These equations extend the degeneracy $\sf n^{\sf gauge}_{\sf g.s.}$ found in 
Subsection \ref{construct_genus} for the Ising (${\sf d}_Q = 2$) lattice gauge theory \cite{local-gauge-Ising}.

\section{$U(1)$ Classical Toric Code Model and its gauge theory limit}
\label{u1sec}
We next turn to a simple $U(1)$ theory

\begin{eqnarray}
\label{fU(1)}
H= - J \sum_{s} \cos(\Phi_{s}) - J'\sum_{p}\cos(\Phi_{p}) ,
\end{eqnarray}
where the ``fluxes''
\begin{eqnarray}
\Phi_{s} = \sum_{i} \theta_{si}, \ \ 
 \Phi_{p} = \sum_{ij \in p} \theta_{ij},
\end{eqnarray}
are, respectively, the sums of the angles on all edges emanating from
site $s$ and the sum of all angles $\theta_{ij}$ on edges that belong to
a plaquette $p$. In the continuum limit (in which the lattice constant
$a$ tends to zero), the $\cos \Phi_{p}$ term may be Taylor expanded as
the flux is small, $\cos \Phi_{p} \approx (1- \frac{1}{2} \Phi_{p}^{2} +
\cdots)$  in the usual way. Then, omitting an irrelevant constant
additive term, the Hamiltonian becomes in the standard manner 
\begin{eqnarray}
H= \frac{1}{2} \int \Phi_{p}^{2}(x) d^{2}x \approx a^{2} \int B_3^2 d^{2}x,
\end{eqnarray}
where $B_3=\partial_{1} A_{2} - \partial_{2} A_{1}$ (with $\vec{A}$ a
vector potential) is the conventional magnetic field along the direction
transverse to the plane where the lattice resides.  In the ${\sf d}_Q
\to \infty$ limit, the $U(1)$ Hamiltonian of Eq. (\ref{fU(1)}) follows
from Eqs. (\ref{sijk}),  (\ref{NewA}), and (\ref{NewB}) where
$\sigma_{ij}=e^{{\rm i}\theta_{ij}}$, and $\theta_{ij}= 2\pi \alpha_{ij}/{\sf
d}_Q$ with $\alpha_{ij}=0,1,\cdots,{\sf d}_Q-1$.  In the ${\sf d}_Q \to
\infty$ limit, the discrete clock symmetry becomes  a continuous
rotational symmetry, $\mathbb{Z}_{{\sf d}_Q} \to U(1)$. Rather
trivially, yet notably, in this limit, the system becomes {\em
gapless}.  Repeating {\it mutatis mutandis} the considerations of Eqs.
(\ref{Deg-Gen}) and (\ref{mindg}), in the continuous large ${\sf d}_Q$
limit, a genus dependent symmetry is naturally associated with the
system degeneracy. Peculiarly, in this limit, similar to Eq.
(\ref{zdq}), a genus dependent
\begin{eqnarray}
\label{Geq}
G= U(1) \otimes U(1) \cdots \otimes U(1)
\end{eqnarray}
symmetry may appear for the Toric $U(1)$ theory of Eq. (\ref{fU(1)}). 
In the limiting case in which the star term does
 not appear in Eq. (\ref{fU(1)}), i.e., that of $J =0$, 
  a symmetry of the type of Eq. (\ref{Geq}) compounds the
 known local $U(1)$ symmetry,
\begin{eqnarray}
 \theta_{ij} \to \theta_{ij} + \phi_{i} - \phi_{j},
\end{eqnarray}
similar to Eq. \eqref{phaseijclock} but with an arbitrary real phase 
$\phi_{i}$ at each lattice vertex (site) $i$.  These
 local symmetries are lifted once the $\cos \Phi_{s}$ term is introduced, 
 as in Eq. (\ref{fU(1)}).  Thus, similar to the Clock gauge theory 
 (whose degeneracy was given by Eqs. (\ref{Min_gauge}), and 
 (\ref{Min_gauge_Con})), this {\it $U(1)$
  lattice gauge theory exhibits a genus dependent
 degeneracy}. 
 
 \begin{figure}[htb]
\centering
\includegraphics[width=0.91 \columnwidth]{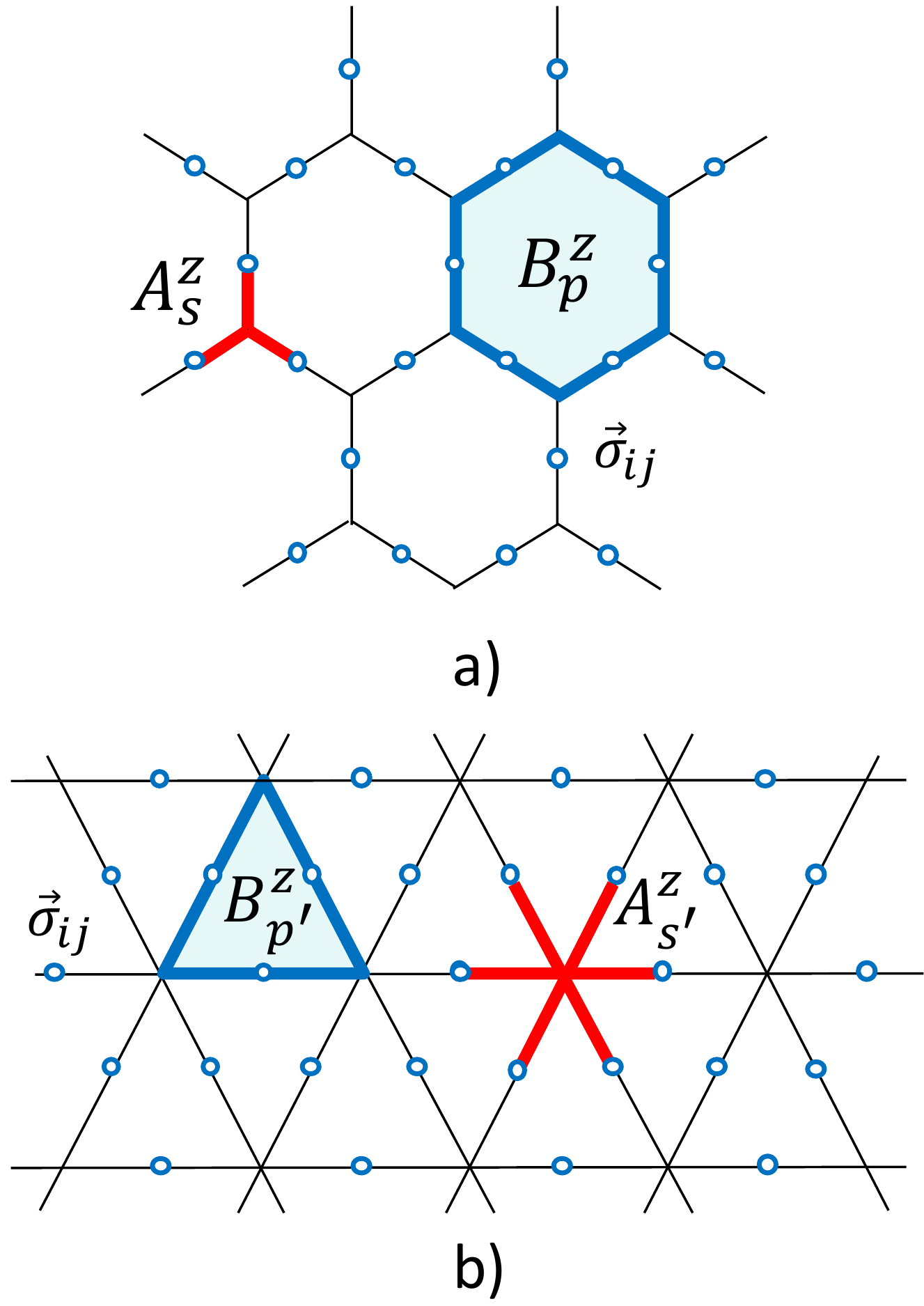}
\caption{ a) Hexagonal lattice and b) Triangular lattice. In panel a)
the star terms $A^z_s$ and plaquette terms $B^z_p$  involve three  and
six spins $S$ (circles) interactions, respectively, while  the opposite
happens in  panel  b). } 
\label{Hex-Tri-1.png}
\end{figure}

\section{Honeycomb and Triangular lattices}
\label{lattice_type}

Thus far, we focused on square lattice realizations of the Ising, clock,
and $U(1)$ theories. For completeness, we now examine other lattice geometries.
Specifically, we study the honeycomb lattice ({\sf H}) and triangular
lattice ({\sf T}) incarnations of our classical theory and determine
their ground state degeneracies.  In Fig. \ref{Hex-Tri-1.png}, $A^z_s$
and $B^z_p$ are defined for each lattice. The Hamiltonians  are given by
\begin{eqnarray}
\label{HTH}
H_{\sf H}&=&- J_h \sum_{s} A^{z}_s -J_h' \sum_{p} B^{z}_p,\nonumber\\
H_{\sf T}&=& - J_t \sum_{s'} A^{z}_{s'} -J'_t \sum_{p'} B^{z}_{p'}.
\end{eqnarray}
Our numerical results are summarized in Table \ref{table4}.
These results are consistent with Eqs. \eqref{Deg-Gen} and \eqref{mindg}.
\begin{figure}[htb]
\centering
\includegraphics[width=0.6 \columnwidth]{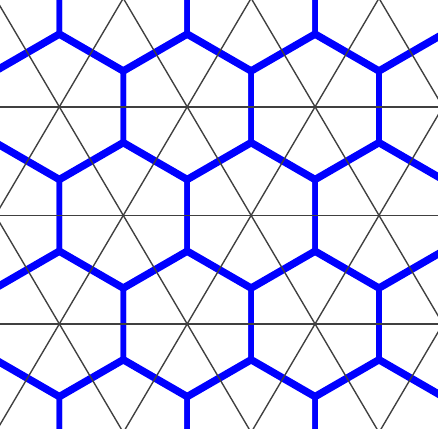}
\caption{By connecting the centers of hexagons in an hexagonal lattice
(thick solid lines), we obtain the corresponding dual lattice which is a
triangular lattice (solid lines). } 
\label{Hex-Tri-2.png}
\end{figure}

\begin{table}[htb]
\centering
\begin{tabular}{|c|c|c|c|c|c|c|c|c|c|}
\cline{1-10} 
$g$ & $E$ & $D^{0}_2$ & $D^{0}_3$& $D^{0}_4$& $D^{0}_5$& $D^{0}_6$ & $D^{0}_7$ & $D^{0}_8$ & $D^{0}_9$  
 \\ \cline{1-10}
\multicolumn{1}{ |c }{\multirow{4}{*}{1} } &
\multicolumn{1}{ |c| } {6} & 8 & 27 & 64 & 125 & 216 & 343 & 512 & 729  \\ \cline{2-10}
\multicolumn{1}{ |c  }{}                        &
\multicolumn{1}{ |c| } {12} & 16 & 27 & & & & & & \\ \cline{2-10}
\multicolumn{1}{ |c  }{}                        &
\multicolumn{1}{ |c| } {18} & 8 & & & & & & &  \\ \cline{2-10}
\multicolumn{1}{ |c  }{}                        &
\multicolumn{1}{ |c| } {24} & 128 &  & & & & & &   \\ \cline{1-10}
\multicolumn{1}{ |c }{\multirow{2}{*}{2} } &
\multicolumn{1}{ |c| } {24} & 128 & &  & & & & & \\ \cline{2-10}
\multicolumn{1}{ |c  }{}                        &
\multicolumn{1}{ |c| } {30} & 64 &  & & & & & &   \\ \cline{1-10}
\end{tabular}
\caption{Computed ground state degeneracy $D^{0}_{\sf M}$
for ${\sf d}_Q={\sf M}$,  for a hexagonal
lattice ($=$ triangular lattice).}
\label{table4}
\end{table}

As is well known, the {\sf H} and {\sf T} lattices are dual lattices
(Fig. \ref{Hex-Tri-2.png}). This duality implies that the classical
Toric Code models of Eq. (\ref{HTH}) yield the same results. From Figs.
\ref{Hex-Tri-1.png} and \ref{Hex-Tri-2.png}, as a consequence  of
duality, what is defined as $A^z_s$ ($B^z_p$) in {\sf H} corresponds to
some $B^{z}_{p'}$ ($A^{z}_{s'}$) in {\sf T}, and vice versa. This
indicates that
\begin{eqnarray}
\label{Duality}
A^{z}_s &\overset{\sf Duality}{\longleftrightarrow}& B^{z}_{p'}, \nonumber\\
A^{z}_{s'} &\overset{\sf Duality}{\longleftrightarrow}& B^{z}_p.
\end{eqnarray}
After this transformation we can rewrite Eqs. (\ref{HTH}) as,
\begin{eqnarray}
\label{DHTH}
H_{\sf H}&=&- J_h \sum_{p'} B^{z}_{p'} -J_h' \sum_{s'} A^{z}_{s'},\nonumber\\
H_{\sf T}&=& - J_t \sum_{p} B^{z}_{p} -J'_t \sum_{s} A^{z}_{s} ,
\end{eqnarray}
and assuming $J_h=J'_t, J_h'=J_t$, it is seen that $H_{\sf H}=H_{\sf T}$.
This simple analysis does not take into account potential boundary terms
that  may appear  in finite lattices, as a result of the duality
transformation. 
 
\section{Other classical models with holographic degeneracy}
\label{MM}

In this section, we dwell on a few more Ising type spin systems,
similar  to Type II commensurate lattice realizations of the classical
Toric Code model (Eq. (\ref{cdeg})),  in which the degeneracy is
holographic, i.e., exponential in the system's boundary. 

\subsection{Potts Compass Model}
\label{potts_s}

\begin{figure}[htb]
\centering
\includegraphics[width=0.9 \columnwidth]{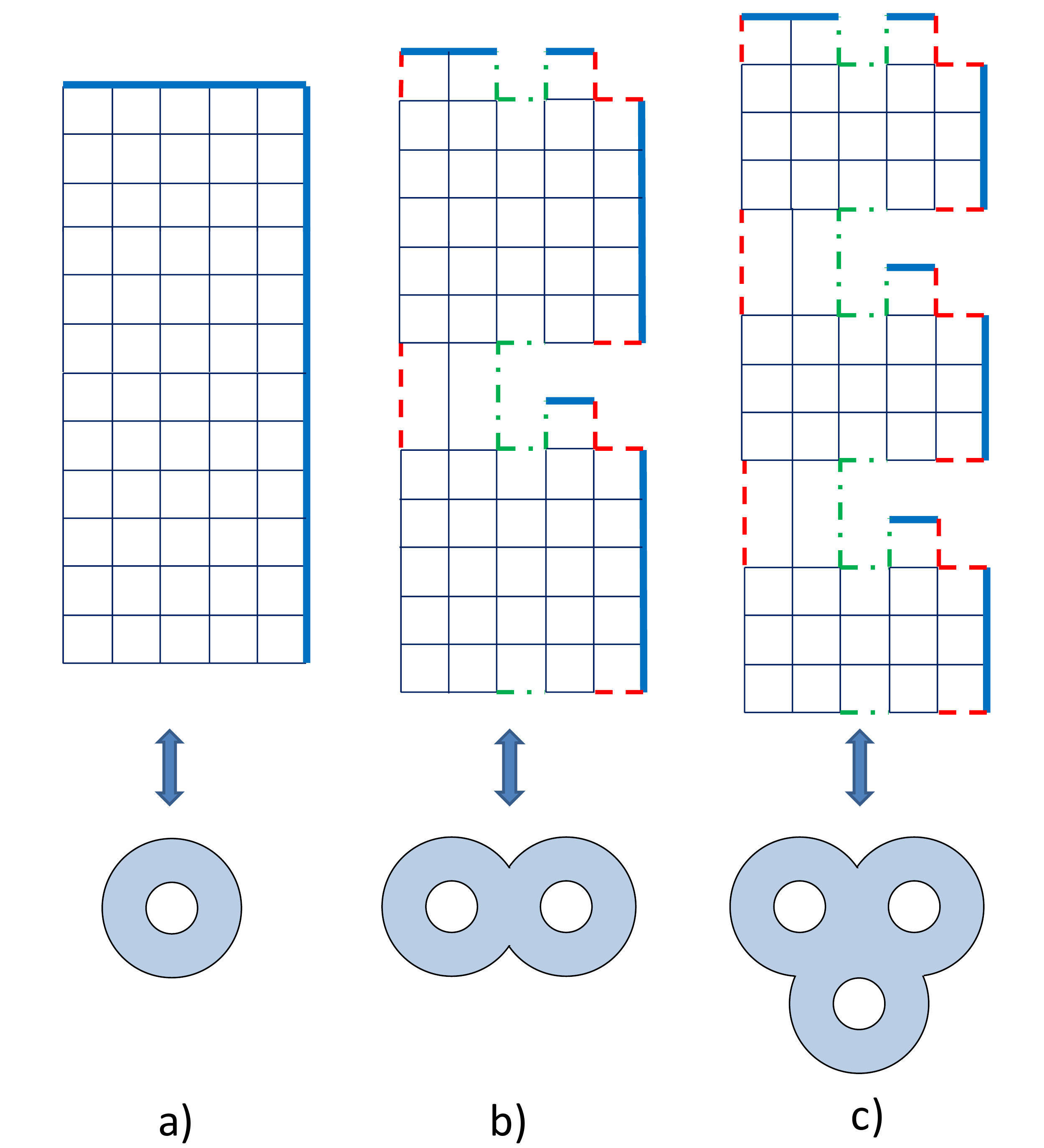}
\caption{Three lattices with different genus numbers and their
corresponding tori below. All have the same total number of spins, $N=60$.
Thick solid (blue) lines represent the boundary and spins are located at the
vertices. We have,
a)  $g=1$ and $L_x=5, L_y=12$.
b)  $g=2$ and $L_x=7, L_y=12$.
c)  $g=3$ and $L_x=9, L_y=12$. }
\label{Genus.png}
\end{figure}

We now discuss a discretized version of the compass model \cite{Compass
Model},  the ``$4$-state Potts compass model'' on an $L_x \times L_y$
square lattice with periodic boundary conditions. The Hamiltonian is
given by,
\begin{eqnarray}
\label{H_P}
 H_{\sf PC}= - \sum_{i,\sigma,\tau} \Big( n_{i\sigma}n_{i+\hat{x},\sigma}
\sigma_{i} \sigma_{i+\hat{x}} + n_{i\tau}n_{i+\hat{y},\tau} \tau_{i}
\tau_{i+\hat{y}} \Big), \nonumber\\
\end{eqnarray}
where at each site (vertex) $i$ there are  two Ising type spins
$\sigma_{i} = \pm 1$, $\tau_{i} = \pm 1$, while the occupation numbers
$n_{i\sigma}=0,1$ and $n_{i\tau} = 1 - n_{i\sigma}$. Then, at each site,
there is either a $\sigma$ or a $\tau$ degree of freedom. The Cartesian
unit vectors  $\hat{x}$ and $\hat{y}$ link neighboring sites of the
square lattice.   Spins of the $\sigma$ type interact along the 
$x$-direction (horizontally) while those of the $\tau$ variety interact
along the $y$-direction (vertically). Minimizing the energy is
equivalent to maximizing the number of products in the summand of Eq.
(\ref{H_P}) that are equal to $+1$. In a configuration in which at all
sites there is a $\sigma$ (and no $\tau$) spin, the system effectively
reduces to that of $L_{y}$ independent Ising chains parallel to the $x$
direction.  For each such chain, there are two ground states: 
$\sigma_i=+1$ or $\sigma_i=-1$ for all lattice sites. As these chains
are independent, there are $2^{L_y}$ ground states. Replacing some sites
with $\tau$ spins some bonds turn into $0$ and energy increases as a
result. Repeating the same procedure where all sites are occupied by
$\tau$ spins, we find out that there are $L_x$ independent vertical
Ising chains and so $2^{L_x}$ states giving the same minimum energy. The
ground state degeneracy of Eq. (\ref{H_P}) is $2^{L_x}+2^{L_y}$.  For a
more general case with genus $g$ (composed of regions $\{a_{j}\}$
connected by bridges $\{b_{j}\}$ (shared by regions $a_{j}$ and
$a_{j+1}$)), the degeneracy again depends on the number of independent
horizontal ($L_y$) and vertical ($L_x$) Ising chains. If each region
$a_j$ is of size $L^{j}_{x} \times L^{j}_{y}$ ($j=1,\cdots,g$) and
$b_j$ ($j=1,\cdots,g-1$) is the number of edges connecting $a_{j}$ 
and $a_{j+1}$, then, the ground state degeneracy will be 
 \begin{eqnarray}
 \label{npc}
 n^{\sf Potts-compass}_{\sf g.s} = 2^{L_x}+ 2^{L_y},
 \end{eqnarray}
 where
\begin{eqnarray}
\label{lylx+}
L_x = \sum^{g}_{j=1} L^{j}_{x} - \sum^{g-1}_{j=1} b_j, ~ L_y = \sum^{g}_{j=1} L^{j}_{y}.
\end{eqnarray}
This degeneracy depends on both the geometry and the topology of the
lattice. We briefly highlight the effects of topology in the degeneracy
of Eqs. (\ref{npc}) and (\ref{lylx+}). Panel a) of  Fig. \ref{Genus.png}
depicts a genus one lattice for which $L_x=5, L_y=12$ and $N=V=60$. By
redefining the way spins are connected and boundary conditions, as we
explained before, we may transform it into, e.g., $g=2,3$ lattices as in
Fig. \ref{Genus.png} (panels b) and c), respectively). Here, one may
readily verify that although $L_y= 12$ and the total number of spins do
not change, $L_x$ varies (increases) as a result of increasing the genus
number.

\subsection{Classical Xu-Moore Model}
\label{xmc}

As discussed earlier, our classical Toric Code model of Eq. (\ref{H'}) 
is identical to the spin (defined on vertices) plaquette model of Eq.
(\ref{HW}). This latter Hamiltonian is, as it turns out,  a particular
limiting case of the so-called ``Xu-Moore model''
\cite{Xu-Moore1,Xu-Moore2}, one in which its transverse field  is set to
zero and the model becomes classical. In its original rendition, this
classical limit of the Xu-Moore model has a degeneracy exponential in
the system's boundary.  This degeneracy appears regardless of the parity
of the system sides. We now discuss how to relate the degeneracy in our
system to that of the classical Xu-Moore model. To achieve this, instead
of applying periodic boundary conditions along the Cartesian directions
as in the classical Toric Code model (i.e. along the solid lines of 
Fig. \ref{XM.png}), we endow the
system with different boundary conditions. Specifically, we examine
instances in which periodic boundary conditions are associated with the
diagonal $x'$ and $y'$ axis ($45^\circ$ angle rotation of the original
square lattice) of Fig. \ref{XM.png}.  A simple calculation then
illustrates that the ground state sector as well as all other energies
have a global degeneracy factor,
\begin{eqnarray}
{\cal{N}}_{\sf global} = 2^{L_{x'} + L_{y'}}.
\end{eqnarray}
where $L_{x'}$ and $L_{y'}$ are defined as in Eq. (\ref{lylx+}) but along the 
diagonal directions (dotted lines in Fig. \ref{XM.png}). A similar
(global) degeneracy appears in the classical 90$^\circ$ orbital compass
model \cite{kitaev_review} (having only nearest neighbor two-spin
interactions) to which the Xu-Moore model is dual.

\subsection{Second and Third nearest neighbor Ising models}
\label{short_Ising}
 
We conclude our discussion of holographic degeneracy in spin models with
 a brief review of an Ising system even simpler than the ones discussed
 above. Specifically,  we may consider an Ising spin system on a square
 lattice with its lattice constant $a$ set to unity when it is embedded on 
 a torus ($g=1$) with periodic boundary conditions along the $x'$
 and $y'$ diagonals with the Hamiltonian 
 \begin{eqnarray}
 H  = \sum_{i,j} ( 2 \delta_{|i-j|,\sqrt{2}} + \delta_{|i-j|,2})
\sigma_{i} \sigma_{j}.
 \end{eqnarray}
Here interactions are anti-ferromagnetic between next-nearest neighbors
($|i-j|=\sqrt{2}$) and next-next-nearest neighbors ($|i-j|=2$). It is
straightforward to demonstrate that this system has a ground state
degeneracy   scales as $2^{L_{x'}} + 2^{L_{y'}}$ where $L_{x',y'}$ are
the lattice sizes along the $x'$ and $y'$ directions \cite{MW4}.
 
\section{Conclusions}
\label{conclusionss}

In this work, we demonstrated that a topological ground state degeneracy
(one depending on the genus number  of the Riemann surface on which the
lattice is embedded) does not imply concurrent topological order (i.e.,
Eq. (\ref{TQO}) is violated and distinct ground states may be told apart
by local measurements). We illustrated this by introducing the classical
Toric Code model (Eq. (\ref{H'}) with $\mu = \nu = z$). As we showed in
some detail, under rather mild conditions  (those pertaining to  ``Type
I'' lattices in the classification of Eq. (\ref{cdeg})), the ground
state degeneracy solely depends on topology.  In these classical
systems, however, the ground states (given by, e.g.,  Eqs. (\ref{gn+})
and  (\ref{ggeneral}) on the torus) are distinguishable by measuring the
pattern of $\sigma^{z}_{ij}$ on a finite number of nearest neighbor
edges; thus, the ground states do not satisfy Eq. (\ref{TQO}) and are,
rather trivially, not topologically ordered. They are Landau ordered 
instead and, most importantly, illustrate that the ground states are 
related by $d=2$ (global) Gauge-like symmetries contrary to the  $d=1$
symmetries of Kitaev's Toric Code model 
\cite{symmetry1,symmetry2,fragility}.  

In the more commensurate Type II lattice realizations of the classical
Toric Code model as well as in a host of other systems,  the ground
state degeneracy is ``holographic''- i.e., exponential in the linear
size of the lattice \cite{MW4,holography}. This classical holographic
effect is  different from more subtle deeper quantum relations, for
entanglement entropies, e.g., \cite{holo1,holo2,holo3}. In all lattices
and topologies, the minimal ground state degeneracy (and that of all
levels in the system) of the classical model is robust and bounded from
below by $4^{g}$ with $g$ the genus number. We find similar genus
dependent  minimal degeneracies in clock and $U(1)$ theories (including
{\it lattice gauge theories}). For completeness, we remark that a
degeneracy of the form $2^{\eta(L)}$ with $\eta$ a quantity bounded from above
by the linear system size (viz., a holographic entropy) also appears in
bona fide topologically ordered systems such as the ``Haah code''
\cite{Haah,Haah1,Haah2}. 

Beyond demonstrating that such degeneracies may arise in classical
theories, we illustrated that these behaviors may arise in rather
canonical clock and $U(1)$ type theories. We provided a simple framework
for studying and  understanding the origin of these ubiquitous
topological and holographic degeneracies. 

We conclude with one last remark. 
Our results for classical systems enable the construction of  simple
{\it quantum models} with ground states that may be told apart
locally (i.e., violating Eq. \eqref{TQO} for topological quantum order) yet, nevertheless, 
exhibit a topological ground state degeneracy). We present one, out of a large number of
possible, routes to write such models exactly. Consider any one of the different theories studied
in our work. Let us denote the classical Hamiltonian associated with any
of these theories by $H_{\sf Classical}$ and corresponding local
observables that may differentiate ground states apart by ${\cal V}$.
One may then apply any product $U$ of local unitary transformations to
both the Hamiltonian and the corresponding  ``order parameter'' local
observable ${\cal V}$. That is, we may consider the ``quantum''
Hamiltonian $H_{\sf Quantum} \equiv U^{\dagger} H_{\sf Classical} U$
and  the corresponding local operator ${\cal V}_{\sf Quantum} \equiv
U^{\dagger} {\cal V} U$. By virtue of the unitary transformation, both
in the ground state sector (as well as at any finite temperature), the
expectation value of the local observable ${\cal V}$ in the classical
system given by $H_{\sf Classical}$ is identical to the expectation
value of the ${\cal V}_{\sf Quantum}$ in the quantum system governed by
$H_{\sf Quantum}$. To be concrete, one may consider, e.g., the
Classical  Toric Code ({\sf CTC}) model. That is, e.g., one may
set $H_{\sf Classical} = H_{\sf CTC}$ that contains only  classical
Ising ($\sigma^x_j$) spins. Next, consider  the unitary operator $U=
\prod_{j \in \Lambda_{+}} \exp[i \frac{\pi}{4} \sigma_{j}^{z}]$ that
effects a $\pi/2$ rotation of all spins at sites $j$ that belong to the
sublattice $\Lambda_{+}$ about the internal $\sigma^z$ axis. (That is,
indeed, $\frac{1}{\sqrt{2}} (1 - i \sigma_{j}^{z}) \sigma_{j}^{x}
\frac{1}{\sqrt{2}}(1+ i \sigma_{j}^{z}) = \sigma_{j}^{y}$.) Thus,
trivially, the resulting Hamiltonian $H_{\sf Quantum}$ contains
non-commuting $\sigma^x$ and $\sigma^y$ and is ``quantum'' (just as
the Kitaev Toric Code model of Section \ref{toric_quantum} 
\cite{Fault-tolerant} that may be mapped to two decoupled 
classical Ising spin chains \cite{symmetry1,symmetry2,fragility})
contains exactly these two quantum spin
components and is ``quantum''). By virtue of the local product nature
of the mapping operator $U$, the classical local observables ${\cal V}$
that we discussed in our paper become now new local observables ${\cal
V}_{\sf Quantum}$ in the quantum model. Thus, putting all of the pieces
together, we may indeed generate quantum models with a topological 
degeneracy in which the ground state may be told apart by local measurements.

\section{Acknowledgment}

This work was partially supported by the National Science Foundation
under NSF Grant No. DMR-1411229 and the Feinberg foundation visiting 
faculty program at the Weizmann Institute.  We are very grateful to a
discussion  with J. Haah in which he explained to us the degeneracies
found in his model \cite{Haah2}. 

\appendix

\section{Canonical Partition function of the Classical Toric Code Model}
\label{Z}

In Type I lattices (and their simplest composites), the canonical
partition function of the classical Toric Code model is given by Eq.
(\ref{ctcz}). The situation is somewhat richer for other lattices.
Below, we briefly write the partition functions for several such finite
size lattices. For simplicity we set $J=J'=1$ and ${\sf d}_Q=2$ in the
classical rendition of Eq. (\ref{H'}) and perform a high temperature
({\sf H-T}) and low temperature ({\sf L-T}) series expansion which is
everywhere convergent for these finite size systems. One can follow a
similar procedure and find the partition functions for ${\sf d}_Q >2$.
We start with {\sf H-T} series expansion,
\begin{eqnarray}
\label{P1}
{\cal Z_{\sf H-T}}&=& \sum_{ \{\sigma \} } e^{ -\beta H^{z,z}}  = \sum_{ \{\sigma \} }
 e^{ \beta \sum_{s} A^{z}_{s} + \beta \sum_{p} B^{z}_{p}}  \\
 &=& \sum_{ \{\sigma \} } \prod_{s} e^{\beta A^{z}_{s}}  \prod_{p}
e^{\beta B^{z}_p} \nonumber\\
 &=& (\cosh\beta)^{V+F} \sum_{ \{\sigma \} } \prod_{s} (1 + T\,  A^{z}_{s} ) 
  \prod_{p} (1+ T \, B^{z}_p ),\nonumber
\end{eqnarray}
where $T= \tanh\beta$ and $\beta = 1/(k_B T)$.

In Eq. (\ref{P1}) after expanding the products, and summing over all
configurations, the only surviving terms are those for which the product
of a subset of $A^z_{s}$'s and $B^z_{p}$'s is equal to $1$ and this
corresponds to one constraint or a product of two or more of them
sharing no star or plaquette operators. Thus, 
\begin{eqnarray}
\label{PF}
{\cal Z_{\sf H-T}}&=& 2^{E} (\cosh\beta)^{V+F} \\
&\times & \Big(1 + \text{terms from constraints on  $A^z_{s}$'s and
$B^z_{p}$'s}  \Big),\nonumber
\end{eqnarray}
where $F$ is the number of faces and $V$ is the number of vertices. The
factor of $2^{E}$ (with $E=N$ the number of spins or lattice edges)
originates from the summation $\sum_{ \{\sigma \} } 1$ (each
$\sigma^z_{ij}$ has two values ($\pm 1$), with $(ij)=1,\cdots,E$). The
sole non-vanishing traces in Eq. (\ref{P1}) originate from the
constraints of Eqs. (\ref{asbp}) and (\ref{wl}) and their higher genus
counterparts. While this procedure trivially gives rise to  the
partition function of Eq. (\ref{ctcz}) for simple lattices, the
additional constraints in other lattices spawn new terms in the
partition functions.
 
In the following we develop the {\sf L-T} series expansion for ${\sf d}_Q=2$. 
From Eq. (\ref{ctcz}),
\begin{eqnarray}
{\cal Z_{\sf L-T}} &=& {\cal{N}}_{\sf global} \sum_{\ell=0} n_{\ell}
e^{-\beta E_{\ell}}\nonumber\\
&=& {\cal{N}}_{\sf global}e^{-\beta E_{0}} \Big( 1 +\sum_{\ell=1}
n_{\ell} e^{-\beta ( E_{\ell}-E_{0})}\Big),
\end{eqnarray}
where $E_{0}$ is the ground state energy and  ${\cal{N}}_{\sf global}$
is the ground state degeneracy. Numerical results illustrate that the
integers $n_{\ell}$ are larger than or equal to $1$.  One can generalize this form
for ${\sf d}_Q>2$
\begin{eqnarray}
{\cal Z_{\sf L-T}}&=& \sum_{\ell=0} D^{\ell}_{{\sf d}_Q} e^{ -\beta E_{\ell}},
\end{eqnarray} 
where $E_{\ell}$ and $D^{\ell}_{{\sf d}_Q}$ indicate energy and
degeneracy of energy level $\ell$ for a given ${\sf d}_Q$, respectively. 

Below is a sample of our numerical results for ${\cal Z}_{\sf H-T}$ and 
${\cal Z}_{\sf L-T}$ of lattices with different sizes, ${\sf d}_Q$'s and
genus numbers ($g=1,2,3$). From ${\cal Z}_{\sf L-T}$, we can easily see
that exited states have a degeneracy ``higher than or equal to'' the
ground state degeneracy ($J=J'$ and  $\beta J={\rm K}$).

\begin{enumerate}[label=(\Roman*)]
\item $\bm{g=1}$:

\begin{enumerate}[label=(\alph*)]

\item $\bm{3\times1, E=6}$:

\begin{enumerate}[label=(\roman*)]
\item \underline{${\sf d}_Q=2$:}

\begin{eqnarray}
{\cal Z_{\sf H-T}}&=& (2\cosh\beta)^{6} \Big(1+ T^6+2 T^3 \Big),\nonumber\\
{\cal Z_{\sf L-T}}&=& 4(e^{6 \text{K}})\Big(1+ 9 e^{-8 \text{K}}+6 e^{-4 \text{K}} \Big).\nonumber
\end{eqnarray}

\item \underline{${\sf d}_Q=3$:}
\begin{eqnarray}
{\cal Z_{\sf H-T}}&=& (3\cosh\beta)^{6} \Big(1+ \frac{T^6}{32}+\frac{3 T^4}{8} \Big) ,\nonumber\\
{\cal Z_{\sf L-T}}&=& 9(e^{6 \text{K}})\Big(1 +10 e^{-9 \text{K}}+12 e^{-\frac{15 \text{K}}{2}}+
36 e^{-6 \text{K}}\nonumber\\
&+&16 e^{-\frac{9 \text{K}}{2}}+6 e^{-3 \text{K}} \Big).\nonumber
\end{eqnarray}

\item \underline{${\sf d}_Q=4$:}
\begin{eqnarray}
{\cal Z_{\sf H-T}}&=& (4\cosh\beta)^{6} \Big(1+ \frac{T^6}{16} \Big) ,\nonumber\\
{\cal Z_{\sf L-T}}&=& 8(e^{6 \text{K}})\Big(1 + e^{-12 \text{K}}+12 e^{-10 \text{K}}+135 e^{-8 \text{K}}\nonumber\\
&+&216 e^{-6 \text{K}}+ 135 e^{-4 \text{K}}+12 e^{-2 \text{K}} \Big).\nonumber
\end{eqnarray}

\item \underline{${\sf d}_Q=5$:}
\begin{eqnarray}
{\cal Z_{\sf H-T}}&=& (5\cosh\beta)^{6} \Big(1+ \frac{T^6}{32} \Big) ,\nonumber\\
{\cal Z_{\sf L-T}}&=& 5(e^{6 \text{K}})\Big( 1 + 90 e^{\left(-\sqrt{5}-5\right) \text{K}}+90 e^{\left(\sqrt{5}-5\right) 
\text{K}}\nonumber\\
&+&240 e^{\left(\frac{1}{4} \left(-\sqrt{5}-1\right)+\sqrt{5}-6\right) \text{K}}+30 e^{\left(\frac{1}{2} 
\left(-\sqrt{5}-1\right)-2\right) \text{K}}\nonumber\\
&+&210 e^{\left(\frac{1}{2} \left(-\sqrt{5}-1\right)+\sqrt{5}-7\right) \text{K}}+12 e^{\left(\frac{5}{4} 
\left(-\sqrt{5}-1\right)-5\right) \text{K}}\nonumber\\
&+&20 e^{\left(\frac{3}{2} \left(-\sqrt{5}-1\right)-6\right) \text{K}}+240 e^{\left(\frac{1}{4} \left(\sqrt{5}-1\right)
-\sqrt{5}-6\right) \text{K}}\nonumber\\
&+&120 e^{\left(\frac{1}{2} \left(-\sqrt{5}-1\right)+\frac{1}{4} \left(\sqrt{5}-1\right)-3\right) \text{K}}\nonumber\\
&+&120 e^{\left(\frac{3}{4} \left(-\sqrt{5}-1\right)+\frac{1}{4} \left(\sqrt{5}-1\right)-4\right) \text{K}}\nonumber\\
&+&60 e^{\left(\frac{5}{4} \left(-\sqrt{5}-1\right)+\frac{1}{4} \left(\sqrt{5}-1\right)-6\right) \text{K}}\nonumber\\
&+&30 e^{\left(\frac{1}{2} \left(\sqrt{5}-1\right)-2\right) \text{K}}\nonumber\\
&+&210 e^{\left(\frac{1}{2} \left(\sqrt{5}-1\right)-\sqrt{5}-7\right) \text{K}}\nonumber\\
&+&120 e^{\left(\frac{1}{4} \left(-\sqrt{5}-1\right)+\frac{1}{2} \left(\sqrt{5}-1\right)-3\right) \text{K}}\nonumber\\
&+&360 e^{\left(\frac{1}{2} \left(-\sqrt{5}-1\right)+\frac{1}{2} \left(\sqrt{5}-1\right)-4\right) \text{K}}\nonumber\\
&+&360 e^{\left(\frac{3}{4} \left(-\sqrt{5}-1\right)+\frac{1}{2} \left(\sqrt{5}-1\right)-5\right) \text{K}}\nonumber\\
&+&120 e^{\left(\frac{1}{4} \left(-\sqrt{5}-1\right)+\frac{3}{4} \left(\sqrt{5}-1\right)-4\right) \text{K}}\nonumber\\
&+&360 e^{\left(\frac{1}{2} \left(-\sqrt{5}-1\right)+\frac{3}{4} \left(\sqrt{5}-1\right)-5\right) \text{K}}\nonumber\\
&+&240 e^{\left(\frac{3}{4} \left(-\sqrt{5}-1\right)+\frac{3}{4} \left(\sqrt{5}-1\right)-6\right) \text{K}}\nonumber\\
&+&12 e^{\left(\frac{5}{4} \left(\sqrt{5}-1\right)-5\right) \text{K}}\nonumber\\
&+&60 e^{\left(\frac{1}{4} \left(-\sqrt{5}-1\right)+\frac{5}{4} \left(\sqrt{5}-1\right)-6\right) \text{K}}\nonumber\\
&+&20 e^{\left(\frac{3}{2} \left(\sqrt{5}-1\right)-6\right) \text{K}} \Big).\nonumber
\end{eqnarray}

\item \underline{${\sf d}_Q=6$:}
\begin{eqnarray}
{\cal Z_{\sf H-T}}&=& (6\cosh\beta)^{6} \Big(1+ \frac{T^6}{32} \Big) ,\nonumber\\
{\cal Z_{\sf L-T}}&=& 36(e^{6 \text{K}})\Big(1 + 6 e^{-11 \text{K}}+12 e^{-10 \text{K}}+24 
e^{-\frac{19 \text{K}}{2}}\nonumber\\
&+&10 e^{-9 \text{K}}+48 e^{-\frac{17 \text{K}}{2}}+165 e^{-8 \text{K}}+12 e^{-\frac{15 \text{K}}{2}}\nonumber\\
&+&192 e^{-7 \text{K}}+168 e^{-\frac{13 \text{K}}{2}}+36 e^{-6 \text{K}}+96 e^{-\frac{11 \text{K}}{2}}\nonumber\\
&+&282 e^{-5 \text{K}}+16 e^{-\frac{9 \text{K}}{2}}+114 e^{-4 \text{K}}+60 e^{-\frac{7 \text{K}}{2}}\nonumber\\
&+&6 e^{-3 \text{K}}+24 e^{-\frac{5 \text{K}}{2}}+24 e^{-2 \text{K}} \Big).\nonumber
\end{eqnarray}

\end{enumerate}
\item $\bm{2\times2, E=8}$:

\begin{enumerate}[label=(\roman*)]
\item \underline{${\sf d}_Q=2$:}

\begin{eqnarray}
{\cal Z_{\sf H-T}}&=& (2\cosh\beta)^{8} \Big(1+ 14 T^{4} + T^{8} \Big),\nonumber\\
{\cal Z_{\sf L-T}}&=& 16(e^{8 \text{K}})\Big(1+e^{-16 \text{K}}+14 e^{-8 \text{K}}\Big).\nonumber
\end{eqnarray}

\item \underline{${\sf d}_Q=3$:}
\begin{eqnarray}
{\cal Z_{\sf H-T}}&=& (3\cosh\beta)^{8} \Big(1+ \frac{3 T^8}{128}+\frac{T^6}{8}+\frac{3 T^4}{4}\Big) ,\nonumber\\
{\cal Z_{\sf L-T}}&=& 27(e^{8 \text{K}})\Big(1+18 e^{-12 \text{K}}+16 e^{-21 \text{K}/2}\nonumber\\
&+&80 e^{-9 \text{K}}+64 e^{-15 \text{K}/2}+56 e^{-6 \text{K}}+8 e^{-3 \text{K}}\Big).\nonumber
\end{eqnarray}

\item \underline{${\sf d}_Q=4$:}
\begin{eqnarray}
{\cal Z_{\sf H-T}}&=& (4\cosh\beta)^{8} \Big(1+ \frac{T^8}{16}+\frac{3 T^4}{4} \Big) ,\nonumber\\
{\cal Z_{\sf L-T}}&=& 128(e^{8 \text{K}})\Big( 1+e^{-16 \text{K}}+44 e^{-12 \text{K}}+64 e^{-10 \text{K}}\nonumber\\
&+&294 e^{-8 \text{K}}+64 e^{-6 \text{K}}+44 e^{-4 \text{K}}\Big).\nonumber
\end{eqnarray}

\end{enumerate}
\item $\bm{4\times1, E=8}$:

\begin{enumerate}[label=(\roman*)]
\item \underline{${\sf d}_Q=2$:}

\begin{eqnarray}
{\cal Z_{\sf H-T}}&=& (2\cosh\beta)^{8} \Big(1+ 2 T^{4} + T^{8} \Big),\nonumber\\
{\cal Z_{\sf L-T}}&=& 4(e^{8 \text{K}})\Big(1+e^{-16 \text{K}}+12 e^{-12 \text{K}}+38 e^{-8 \text{K}}\nonumber\\
&+&12 e^{-4 \text{K}}\Big).\nonumber
\end{eqnarray}

\item \underline{${\sf d}_Q=3$:}
\begin{eqnarray}
{\cal Z_{\sf H-T}}&=& (3\cosh\beta)^{8} \Big(1+ \frac{T^8}{128}\Big) ,\nonumber\\
{\cal Z_{\sf L-T}}&=& 3(e^{8 \text{K}})\Big(1+86 e^{-12 \text{K}}+336 e^{-\frac{21 \text{K}}{2}}\nonumber\\
&+&616 e^{-9 \text{K}}+560 e^{-\frac{15 \text{K}}{2}}420 e^{-6 \text{K}}+112 e^{-\frac{9 \text{K}}{2}}\nonumber\\
&+&56 e^{-3 \text{K}} \Big).\nonumber
\end{eqnarray}

\item \underline{${\sf d}_Q=4$:}
\begin{eqnarray}
{\cal Z_{\sf H-T}}&=& (4\cosh\beta)^{8} \Big( 1+ \frac{T^8}{64} \Big) ,\nonumber\\
{\cal Z_{\sf L-T}}&=& 16(e^{8 \text{K}})\Big(1+ e^{-16 \text{K}}+8 e^{-14 \text{K}}+252 e^{-12 \text{K}}\nonumber\\
&+&952 e^{-10 \text{K}}+1670 e^{-8 \text{K}}+952 e^{-6 \text{K}}\nonumber\\
&+&252 e^{-4 \text{K}}+8 e^{-2 \text{K}}\Big).\nonumber
\end{eqnarray}

\end{enumerate}
\item $\bm{3\times2, E=12}$:

\begin{enumerate}[label=(\roman*)]
\item \underline{${\sf d}_Q=2$:}

\begin{eqnarray}
{\cal Z_{\sf H-T}}&=& (2\cosh\beta)^{12} \Big( 1+ 2 T^6+T^{12} \Big),\nonumber\\
{\cal Z_{\sf L-T}}&=& 4(e^{12 \text{K}})\Big(1+ e^{-24 \text{K}}+30 e^{-20 \text{K}}\nonumber\\
&+&255 e^{-16 \text{K}}+452 e^{-12 \text{K}}+255 e^{-8 \text{K}}+30 e^{-4 \text{K}}\Big).\nonumber
\end{eqnarray}

\item \underline{${\sf d}_Q=3$:}
\begin{eqnarray}
{\cal Z_{\sf H-T}}&=& (3\cosh\beta)^{12} \Big(1+ \frac{T^{12}}{2048}+\frac{3 T^8}{128} \Big),\nonumber\\
{\cal Z_{\sf L-T}}&=& 9(e^{12 \text{K}})\Big(1+ 466 e^{-18 \text{K}}+2664 e^{-\frac{33 \text{K}}{2}}\nonumber\\
&+&7668 e^{-15 \text{K}}+12344 e^{-\frac{27 \text{K}}{2}}+14148 e^{-12 \text{K}}\nonumber\\
&+&11232 e^{-\frac{21 \text{K}}{2}}+6720 e^{-9 \text{K}}+2592 e^{-\frac{15 \text{K}}{2}}\nonumber\\
&+&1026 e^{-6 \text{K}}+152 e^{-\frac{9 \text{K}}{2}}+36 e^{-3 \text{K}}\Big).\nonumber
\end{eqnarray}

\end{enumerate}
\item $\bm{4\times2, E=16}$:

\begin{enumerate}[label=(\roman*)]
\item \underline{${\sf d}_Q=2$:}

\begin{eqnarray}
{\cal Z_{\sf H-T}}&=& (2\cosh\beta)^{16} \Big(1+ T^{16}+14 T^8 \Big),\nonumber\\
{\cal Z_{\sf L-T}}&=& 16(e^{16 \text{K}})\Big(1 + e^{-32 \text{K}}+8 e^{-28 \text{K}}+252 e^{-24 \text{K}}\nonumber\\
&+&952 e^{-20 \text{K}}+1670 e^{-16 \text{K}}+952 e^{-12 \text{K}}\nonumber\\
&+&252 e^{-8 \text{K}}+8 e^{-4 \text{K}} \Big).\nonumber
\end{eqnarray}

\end{enumerate}
\item $\bm{3\times3, E=18}$:

\begin{enumerate}[label=(\roman*)]
\item \underline{${\sf d}_Q=2$:}

\begin{eqnarray}
{\cal Z_{\sf H-T}}&=& (2\cosh\beta)^{18} \Big(1+ T^{18}+6 T^{12}+9 T^{10}\nonumber\\
&+&32 T^9+9 T^8+6 T^6 \Big),\nonumber\\
{\cal Z_{\sf L-T}}&=& 64(e^{18 \text{K}})\Big(1+9 e^{-32 \text{K}}+72 e^{-28 \text{K}}+636 e^{-24 \text{K}}\nonumber\\
&+&1296 e^{-20 \text{K}}+1422 e^{-16 \text{K}}+552 e^{-12 \text{K}}\nonumber\\
&+&108 e^{-8 \text{K}} \Big).\nonumber
\end{eqnarray}

\end{enumerate}
\end{enumerate}

\item $\bm{g=2}$
\begin{enumerate}[label=(\alph*)]

\item $\bm{2\times1 + 2\times1 ,E=8}$:

\begin{enumerate}[label=(\roman*)]
\item \underline{${\sf d}_Q=2$:}
\begin{eqnarray}
{\cal Z_{\sf H-T}}&=& 2^{8}(\cosh\beta)^{6} \Big(1+ T^6+T^4+T^2 \Big),\nonumber\\
{\cal Z_{\sf L-T}}&=& 16(e^{6 \text{K}})\Big(1+ e^{-12 \text{K}}+7 e^{-8 \text{K}}+7 e^{-4 \text{K}} \Big).\nonumber
\end{eqnarray}

\item \underline{${\sf d}_Q=3$:}
\begin{eqnarray}
{\cal Z_{\sf H-T}}&=&  3^{8}(\cosh\beta)^{6} \Big(1+ \frac{T^6}{32} \Big),\nonumber\\
{\cal Z_{\sf L-T}}&=& 27(e^{6 \text{K}})\Big(1 +22 e^{-9 \text{K}}+60 e^{-\frac{15 \text{K}}{2}}\nonumber\\
&+&90 e^{-6 \text{K}}+40 e^{-\frac{9 \text{K}}{2}}+30 e^{-3 \text{K}} \Big).\nonumber
\end{eqnarray}

\item \underline{${\sf d}_Q=4$:}
\begin{eqnarray}
{\cal Z_{\sf H-T}}&=&  4^{8}(\cosh\beta)^{6} \Big(1+\frac{T^6}{16} \Big),\nonumber\\
{\cal Z_{\sf L-T}}&=& 256(e^{6 \text{K}})\Big(1 +e^{-12 \text{K}}+4 e^{-10 \text{K}}+71 e^{-8 \text{K}}\nonumber\\
&+&104 e^{-6 \text{K}}+71 e^{-4 \text{K}}+4 e^{-2 \text{K}} \Big).\nonumber
\end{eqnarray}

\end{enumerate}
\item $\bm{3\times1 + 3\times1 (b_1=1), E=12}$:

\begin{enumerate}[label=(\roman*)]
\item \underline{${\sf d}_Q=2$:}

\begin{eqnarray}
{\cal Z_{\sf H-T}}&=&  2^{12}(\cosh\beta)^{10} \Big(1+ T^{10}+T^6+T^4 \Big),\nonumber\\
{\cal Z_{\sf L-T}}&=& 16(e^{10 \text{K}})\Big(1 + e^{-20 \text{K}}+21 e^{-16 \text{K}}\nonumber\\
&+&106 e^{-12 \text{K}}+106 e^{-8 \text{K}}+21 e^{-4 \text{K}} \Big).\nonumber
\end{eqnarray}

\item \underline{${\sf d}_Q=3$:}

\begin{eqnarray}
{\cal Z_{\sf H-T}}&=&  3^{12}(\cosh\beta)^{10} \Big(1+ \frac{T^{10}}{512}+\frac{T^7}{32}+
\frac{T^6}{32} \Big),\nonumber\\
{\cal Z_{\sf L-T}}&=& 81(e^{10 \text{K}})\Big( 1+ 114 e^{-15 \text{K}}+572 e^{-\frac{27 \text{K}}{2}}+
1266 e^{-12 \text{K}}\nonumber\\
&+&1716 e^{-\frac{21 \text{K}}{2}}+1530 e^{-9 \text{K}}+816 e^{-\frac{15 \text{K}}{2}}\nonumber\\
&+&438 e^{-6 \text{K}}+84 e^{-\frac{9 \text{K}}{2}}+24 e^{-3 \text{K}} \Big).\nonumber
\end{eqnarray}

\end{enumerate}
\item $\bm{3\times1 + 3\times1 (b_1=2), E=12}$:

\begin{enumerate}[label=(\roman*)]
\item \underline{${\sf d}_Q=2$:}

\begin{eqnarray}
{\cal Z_{\sf H-T}}&=& 2^{12}(\cosh\beta)^{10} \Big(1+ T^{10}+T^6+4 T^5+T^4 \Big),\nonumber\\
{\cal Z_{\sf L-T}}&=& 32(e^{10 \text{K}})\Big(1 + 13 e^{-16 \text{K}}+48 e^{-12 \text{K}}+58 e^{-8 \text{K}}\nonumber\\
&+&8 e^{-4 \text{K}} \Big).\nonumber
\end{eqnarray}

\item \underline{${\sf d}_Q=3$:}

\begin{eqnarray}
{\cal Z_{\sf H-T}}&=& 3^{12}(\cosh\beta)^{10} \Big(1+ \frac{T^{10}}{512}+\frac{T^7}{32}+\frac{T^6}{32} \Big),\nonumber\\
{\cal Z_{\sf L-T}}&=& 81(e^{10 \text{K}})\Big( 1 +114 e^{-15 \text{K}}+572 e^{-\frac{27 \text{K}}{2}}+1266 
e^{-12 \text{K}}\nonumber\\
&+&1716 e^{-\frac{21 \text{K}}{2}}+1530 e^{-9 \text{K}}+816 e^{-\frac{15 \text{K}}{2}}+438 e^{-6 \text{K}}\nonumber\\
&+&84 e^{-\frac{9 \text{K}}{2}}+24 e^{-3 \text{K}} \Big).\nonumber
\end{eqnarray}

\end{enumerate}
\item $\bm{2\times2 + 2\times1 , E=12}$:

\begin{enumerate}[label=(\roman*)]
\item \underline{${\sf d}_Q=2$:}

\begin{eqnarray}
{\cal Z_{\sf H-T}}&=& 2^{12}(\cosh\beta)^{10} \Big(1+ T^{10}+3 T^6+3 T^4 \Big),\nonumber\\
{\cal Z_{\sf L-T}}&=& 32(e^{10 \text{K}})\Big( 1 +e^{-20 \text{K}}+9 e^{-16 \text{K}}+54 e^{-12 \text{K}}\nonumber\\
&+&54 e^{-8 \text{K}}+9 e^{-4 \text{K}} \Big).\nonumber
\end{eqnarray}

\item \underline{${\sf d}_Q=3$:}

\begin{eqnarray}
{\cal Z_{\sf H-T}}&=& 3^{12}(\cosh\beta)^{10} \Big(1+ \frac{T^{10}}{512}\Big),\nonumber\\
{\cal Z_{\sf L-T}}&=& 27(e^{10 \text{K}})\Big( 1 +342 e^{-15 \text{K}}+1700 e^{-\frac{27 \text{K}}{2}}\nonumber\\
&+&3870 e^{-12 \text{K}}+5040 e^{-\frac{21 \text{K}}{2}}+4620 e^{-9 \text{K}}\nonumber\\
&+&2520 e^{-\frac{15 \text{K}}{2}}+1260 e^{-6 \text{K}}+240 e^{-\frac{9 \text{K}}{2}}+90 e^{-3 \text{K}} \Big).\nonumber
\end{eqnarray}

\end{enumerate}
\end{enumerate}

\item $\bm{g=3}$:
\begin{enumerate}[label=(\alph*)]

\item $\bm{2\times1 + 2\times1 + 2\times1 ,E=12}$:

\begin{enumerate}[label=(\roman*)]
\item \underline{${\sf d}_Q=2$:}
\begin{eqnarray}
{\cal Z_{\sf H-T}}&=& 2^{12}(\cosh\beta)^{8} \Big(1+ T^8+T^6+T^2 \Big),\nonumber\\
{\cal Z_{\sf L-T}}&=& 64(e^{8 \text{K}})\Big(1 +e^{-16 \text{K}}+16 e^{-12 \text{K}}\nonumber\\
&+&30 e^{-8 \text{K}}+16 e^{-4 \text{K}} \Big).\nonumber
\end{eqnarray}

\item \underline{${\sf d}_Q=3$:}
\begin{eqnarray}
{\cal Z_{\sf H-T}}&=& 3^{12}(\cosh\beta)^{8} \Big(1+\frac{T^8}{128} \Big),\nonumber\\
{\cal Z_{\sf L-T}}&=& 243(e^{8 \text{K}})\Big(1 +86 e^{-12 \text{K}}+336 e^{-\frac{21 \text{K}}{2}}+
616 e^{-9 \text{K}}\nonumber\\
&+&560 e^{-\frac{15 \text{K}}{2}}+420 e^{-6 \text{K}}+112 e^{-\frac{9 \text{K}}{2}}+56 e^{-3 \text{K}} \Big).\nonumber
\end{eqnarray}

\end{enumerate}
\end{enumerate}
\end{enumerate}


\begin{thebibliography}{28}
\bibitem{Haah2} 
Specifically, on a cube of size $L \times L \times L$ endowed with
periodic boundary conditions, the degeneracy of Haah's Code of
\cite{Haah,Haah1} is equal to $2^{\eta(L)}$ with $\eta$ set by the degree (in
$t$) of the polynomial that constitutes the greatest common divisor of
the below three polynomials,
\begin{eqnarray}
\frac{\eta+2}{4} = \deg \gcd \{1+(1+t)^L, 1+(1+ \omega t)^L, \nonumber
\\  1+(1+\omega^2 t)^L\},
\end{eqnarray}
where $\omega = e^{{\rm i} 2 \pi/3}$. This degeneracy varies between $4$
and an exponential in $L$ and can  exhibit capricious jumps as the
lattice size is varied. For instance, (i) when $L=2^{n}$ (with $n$ a
natural number) the degeneracy is $2^{4L-2}$ while (ii) when
$L=2^{n}+1$, the degeneracy is four \cite{Haah,Haah1}.
\bibitem{local-gauge-Ising} 
In any particular fixed gauge,  the degeneracy is $4^{g}$
(see, e.g., Eq. (187) of \cite{symmetry1} for a derivation).
The factor of $2^{\sf N_{site}-1}$ is the number of independent 
gauge fixes as can be seen as follows.
The operators $A^{\mu=x}$ of Eq. (\ref{AB}) connect one gauge fix to another. 
There are $\sf N_{site}$ such operators, and they  are not independent of each other. Specifically, the local 
gauge symmetry operators $A^{\mu=x}$ adhere to the single global 
constraint of Eq. (\ref{ABC}).
\end{thebibliography}
\end{document}